\documentclass[12pt, 1p, a4paper,  authoryear]{elsarticle}
\pdfoutput=1
\usepackage[english]{babel}
\usepackage{amsmath}
\usepackage{amssymb}%
\usepackage{mathrsfs}
\usepackage{amsbsy}
\usepackage{amsthm}
\usepackage{geometry}

\usepackage{natbib}
\usepackage{float}

\usepackage{enumerate}
\usepackage{enumitem}
\usepackage{booktabs}
\usepackage{float}
\usepackage{relsize}
\usepackage{graphics}
\usepackage{longtable}

\usepackage{algorithm}
\usepackage{algpseudocode}
\usepackage{algorithmicx}
\usepackage[]{graphicx}

\usepackage{accents} 

\usepackage[mathscr]{euscript}
\usepackage[table]{xcolor}
\usepackage[flushleft]{threeparttable}

\usepackage{algorithm}
\usepackage{algpseudocode}
\usepackage{algorithmicx}
\usepackage{url}
\usepackage{array}
\usepackage{bookmark}

\usepackage{caption}
\usepackage{subcaption}
\usepackage[modulo]{lineno}

\usepackage[bibliography = common]{apxproof}
\newtheoremrep{thm}[theorem]{Theorem}
\newtheoremrep{lmm}[lemma]{Lemma}
\newtheoremrep{prp}[proposition]{Proposition}[]

\graphicspath{{Figures/}}

\newcommand{\trasps}[1]{^{\raisebox{\depth}{$\scriptstyle{\intercal}$}}_{#1}} 
\newcommand{\traspd}[1]{^{\raisebox{-1.0mm}{$\scriptstyle{\intercal}$}}_{#1}} 

\newcommand{\matinvj}[1]{^{\raisebox{-1.0mm}{$\scriptstyle{{-1}}$}}_{#1}} 

\makeatletter
\newcommand*{\trasp}{%
  {\mathpalette\@trasp{}}%
}
\newcommand*{\@trasp}[2]{%
  ^{\raisebox{\depth}{$\m@th#1\scriptstyle{\intercal}$}}
}
\makeatother

\makeatletter
\newcommand*{\traspj}{%
  {\mathpalette\@traspj{}}%
}
\newcommand*{\@traspj}[2]{%
  ^{\raisebox{\depth}{$\m@th#1\scriptstyle{\intercal}$}}_j
}
\makeatother

\newcommand{\ba}{\mathbf{a}}

\newcommand{\bb}{{\mathbf{b}}}
\newcommand{\be}{{\mathbf{e}}}

\newcommand{\bp}{{\mathbf{p}}}
\newcommand{\bt}{{\mathbf{t}}}

\newcommand{\bx}{{\mathbf{x}}}

\newcommand{\br}{{\mathbf{r}}}

\newcommand{\bv}{{\mathbf{v}}}
\newcommand{\bX}{{\mathbf{X}}}

\newcommand{\bA}{{\mathbf{A}}}

\newcommand{\bC}{{\mathbf{C}}}

\newcommand{\bE}{{\mathbf{E}}}

\newcommand{\bT}{{\mathbf{T}}}

\newcommand{\bR}{{\mathbf{R}}}
\newcommand{\bw}{{\mathbf{w}}}
\newcommand{\bS}{{\mathbf{S}}}

\newcommand{\bQ}{{\mathbf{Q}}}

\newcommand{\bQj}{{\mathbf{Q}_j}}

\newcommand{\btj}{{\mathbf{t}_j}}
\newcommand{\btjT}{{\mathbf{t}\trasps{j}}}
\newcommand{\baj}{{\mathbf{a}_j}}
\newcommand{\bajT}{{\mathbf{a}\trasps{j}}}

\newcommand{\brj}{{\mathbf{r}_j}}
\newcommand{\bwj}{{\mathbf{w}_j}}

\newcommand{\dX}{{\overset{\boldsymbol{.}}{\mathbf{X}}}}

\newcommand{\bdS}{{\overset{\boldsymbol{.}}{\mathbf{S}}}}

\newcommand{\dSj}{{\overset{\boldsymbol{.}}{\mathbf{S}}_j}}
\newcommand{\dSjinv}{{\overset{\boldsymbol{.}}{\mathbf{S}}\matinvj{j}}}

\newcommand{\dXj}{{\overset{\boldsymbol{.}}{\mathbf{X}}_j}}
\newcommand{\dXjT}{{\overset{\boldsymbol{.}}{\mathbf{X}}\traspd{j}}}

\newcommand{\da}{{\overset{\boldsymbol{.}}{\mathbf{a}}}}
\newcommand{\daj}{{\overset{\boldsymbol{.}}{\mathbf{a}}_j}}
\newcommand{\dajT}{{\overset{\boldsymbol{.}}{\mathbf{a}}\traspd{j}}}

\newcommand{\vexp}{\text{vexp}}

\usepackage[printwatermark,disablegeometry]{xwatermark}
\newwatermark[pages=1-35,color=red!25,angle=45,scale = 3,xpos= 0,ypos=0]{Submission}

\renewcommand{\baselinestretch}{\bsln}

\newcommand{\blind}{0}

\begin{document}
\begin{frontmatter}
\title{Sparse Principal Components Analysis: a Tutorial}
\if0\blind{
\author[gmm]{Giovanni Maria Merola}
\ead{giovanni.merola@xjtlu.edu.cn}
\address[gmm]{Xi'an Jiaotong-Liverpool University,
Department of Mathematical Sciences. 111 Ren’ai Road,
Suzhou Industrial Park, Jiangsu Province, PRC
215123.}
}\fi
\begin{abstract}
The topic of this tutorial is Least Squares Sparse Principal Components Analysis (LS SPCA) which is a simple method for computing approximated Principal Components which are combinations of only a few of the observed variables.
Analogously to Principal Components, these components are uncorrelated and sequentially best approximate the dataset. The derivation of LS SPCA is intuitive for anyone familiar with linear regression. Since LS SPCA is based on a different optimality from other SPCA methods
and does not suffer from some serious drawbacks of .
I will demonstrate on two datasets how useful and parsimonious sparse PCs can be computed. An R package for computing LS SPCA is available for download.
\end{abstract}
\end{frontmatter}%

\section{Introduction}
Principal component analysis (PCA) is one of the oldest and most popular methods used to analyze  multivariate data. PCA owes its popularity to being a simple yet useful method that can be applied under generic assumptions on the distribution of observed data. It is included in every book on multivariate analysis and implemented in virtually all statistical analysis packages.

PCA produces linear combinations (weighted sums and differences) of the observed variables, called principal components (PCs). The PCs are mutually uncorrelated and sequentially best approximate the data.

Often, analysts would like to interpret the PCs as meaningful combinations of a few key variables. This can be difficult to do because the PCs are combinations of all the observed variables.
For example, the first PC of the results of 12 ability tests on a sample of students\footnote{This example uses the Students' Ability dataset which will be considered in the examples in Section \ref{sec:ferdata}.} is equal to
\begin{align*}
  & 3.7\% visual + 1.3\% cubes + 3.5\% flags + 1.3\% paragraph + 2.1\% sentence + \\
  &2.8\% wordm + 16.1\% addition + 16.1\% counting + 37.7\% straight + 7.8\% deduct + \\
  &2.7\% numeric + 4.8\% series.
\end{align*}
It is difficult to describe this linear combination with one sentence, or even with two.
Here I show the coefficients scaled to percentage \emph{contributions} (so, the sum of the absolute values is equal to one) because these are easy to interpret. I use the term \emph{loadings} for the coefficients scaled to unit sum of squares
\footnote{The term loadings for the coefficients has been introduced in recent literature. I will follow it even though it creates ambiguity with the jargon used in different dimensionality reduction methods.}.

The most commonly used method to simplify the interpretation of the PCs, called \emph{thresholding}, is to consider only the loadings with (absolute) value of larger than a threshold value.
For example the PC in the above example could be interpreted  as
\[
50\% {straight} + 25\% {addition} + 25\%  {counting}.
\]
Such simplification is said to have \emph{cardinality} equal to three, because only three loadings are not equal to zero.

Thresholding is considered a misleading practice \citep{jol00}, for several reasons, including that: the loadings selected would be different if the others were really equal to zero; larger loadings often correspond to highly correlated variables; and the choice of the threshold values is subjective \citep[see][for a discussion on thresholding the PCs]{mer19rot}.

In the last 20 years, a large number of sparse PCA (SPCA) methods have been proposed to replace thresholding. These methods, to which I refer as ``conventional'', produce components with some genuinely zero loadings, called sparse principal components (SPCs). SPCA seems to have become popular mainly within the machine learning community, maybe because the methods are usually presented as intimidating optimization ``black-boxes'' and require the tuning of obscure parameters. Another reason could be that the SPCs computed are the PCs of subsets of highly correlated variables \citep{mog}, which are not orthogonal and do not approximate well neither the PCs nor the data. Components' orthogonality is extremely important because it allows to interpret each one of them irrespectively of the others. Instead, when the components are correlated a change in one presumes a change in the others.

In \cite{mer} I proposed least squares SPCA (LS SPCA) which is derived by simply adding a sparsity requirement to PCA. Hence, it computes orthogonal SPCs that sequentially explain the most possible variance of the data (considering the constraints), just like the PCs. LS SPCA does not suffer from the same drawbacks as the other SPCA methods. It is also easy to understand and to compute, as I will show in this tutorial.

Since the goal is to explain as much variance of the data as possible with sparse components, and the PCs explain the most, approximating the PCs with SPCs also produces good solutions. So, I proposed, Projection SPCA (PSPCA) to compute suboptimal SPCs by simply projecting (by linear regression) the PCs on a subset of variables, in  \cite{mer19}

\begin{figure}[H]
  \centering
  \includegraphics[width=0.75\textwidth]{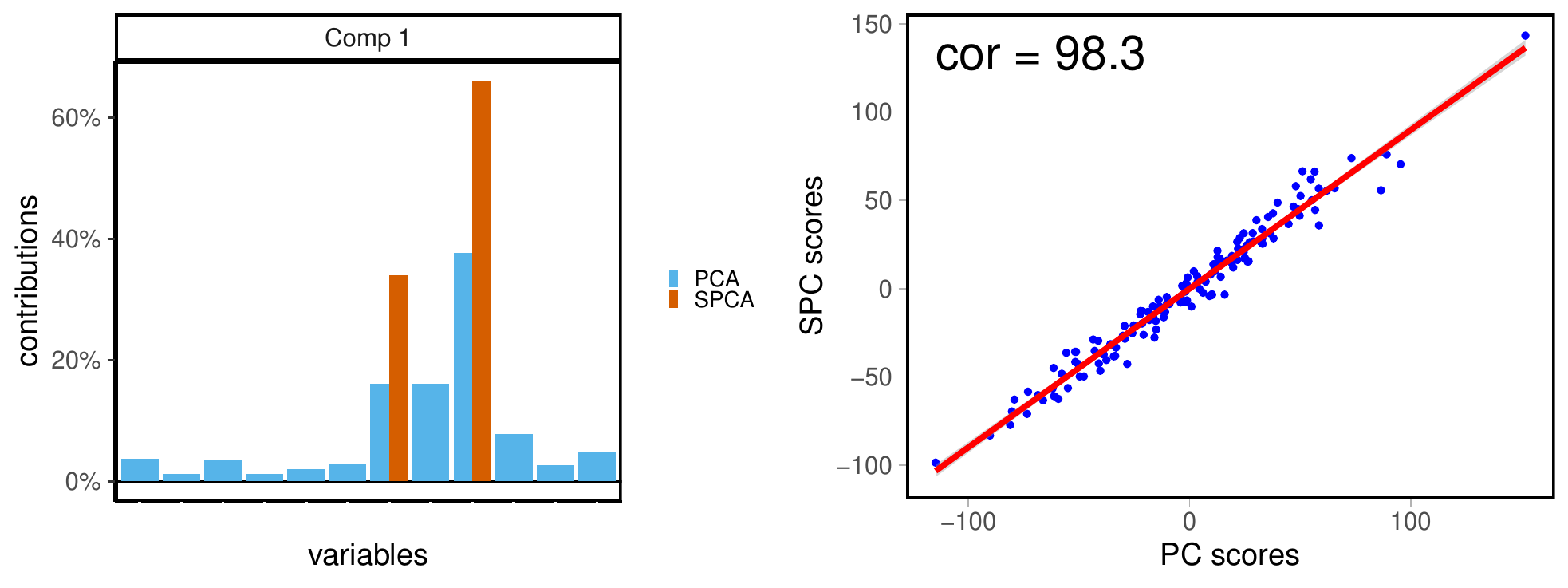}
  \caption{Loadings of the first LSSPC and its scores against the standard PC.}
  \label{fig:intro_loads}
\end{figure}

As an example,  LSSPCs would produce a sparse approximation to the PC mentioned above equal to
\[
34\%addition + 66\%straight.
\]
These contributions are shown in the plot on the left in Figure \ref{fig:intro_loads} with the PC contributions. The scatter plot of their scores (their values), on the right of the same picture, shows how the SPC is almost perfectly correlated with the PC even though it is a combination of only two of the variables.
I hope that this example is useful to show how powerful LS SPCA can be in simplifying the PCs.
In the following I will first show how LS SPCA is derived and its differences with conventional SPCA methods. In Section \ref{sec:example} I will give examples on two data sets. Following this,  in Section \ref{sec:compDet}, I will give the basic computational formulae. Lastly, I will give
some concluding remarks. An R \cite{R} package is available for download.
\section{Derivation of the SPCA solutions}\label{sec:outline}
In this section I will try to keep the mathematical details as simple as possible but some linear algebra is necessary. Further details and proofs can be found in my papers referenced below. This section can be omitted by readers only interested in practical applications.

In the following I denote by $\bX$ an $n\times p$ matrix containing $n$ observations on $p$ variables centered to zero mean by subtracting their average from each observation.
\subsection{PCA}
I assume that readers are familiar with the basics of PCA, otherwise, PCA is presented in monographs \citep[for example,][]{jol, jac} and in every book on multivariate analysis \citep[among others]{ize08, ada}. Here I give just the concepts necessary to understand SPCA. The PCs are linear combinations of the variables, denoted by $\bp_j = \sum_{k = 1}^{p} \bx_j v_{kj}$, where the coefficients  $v_{kj}$ are the loadings of the $k$-th variable in the $j$-th PC.
Note that bold uppercase letters denote matrices and bold lower case letters vectors.

PCA is optimal with respect to several criteria \footnote{PCA can be derived as the solution to many problems. For example, as Karhunen–Loeve transform and empirical orthogonal functions. Essentially, Pearson's definition of PCA is equivalent to the singular value decomposition \citep{eck}. \cite{rao64} gives a brilliant review of various optimal property of PCA.}.
Sparse PCA methods are derived from either of the two definitions of PCA most commonly used in Statistics: those of \cite{pea} and \cite{hot}.

\cite{pea} defines the PCs as the mutually orthogonal linear combinations of the variables that sequentially yield the best least squares approximation of the data.
Hence, the first $d < p$ PCs are the least squares solutions of the multivariate regression model
\begin{equation}\label{eq:pearpca1}
  \bX = \bp_{1}\bb\trasps{1} + \cdots + \bp_d\bb\trasps{d} + \bE,\, \text{subject to } \bp\trasps{j}\bp_k = 0\, \text{if } k > j,
\end{equation}
where $\bb_k$ denotes a vector of regression coefficients, the superscript $^{\scriptstyle{\intercal}}$ denotes transposition and $\bE$ is a matrix of residuals.
The difference with standard regression is that we only know the regressors up to being a linear combination of the responses, but the coefficient vectors $\bb_j$ are still given by the standard LS formula. The least squares solution (that minimizes the sum of the squared residuals, $\sum_{i = 1}^n\sum_{j =1}^p e_{ij}^2$) can be found with simple linear algebra. The loadings $\bv_k = (v_{1k}, \ldots, v_{pk})\trasp$ are the eigenvectors of the covariance matrix $\bS \propto \bX\trasp\bX$ (PCA is invariant to changes of scale of the covariance matrix, so it is enough to define it up to a scalar constant). Hence, the loadings must sequentially  maximize the equation
\begin{equation}\label{eq:pcaeigenvecs}
  \frac{\bv\trasps{k}\bS\bv_k}{\bv\trasps{k}\bv_k} = \lambda_k,
\end{equation}
where $\lambda_k$ is the largest possible eigenvalue under the (orthogonality) constraints $\bv\trasps{k}\bS\bv_j = 0$ for all $j < k$. The eigenvalues are taken in nondecreasing order, so that $\lambda_k \geq \lambda_i$ if $k < i$.

The \emph{variance explained} by a PC is equal to the variance (the sum of squares) of the approximation
$\bp_k\bb\trasps{k}$ in equation \ref{eq:pearpca1}. It is easy to prove  that this is equal to the corresponding  eigenvalue, $\lambda_j$, so that $\lambda_1 + \cdots \lambda_d$ is the cumulative variance explained by the first $d$ PCs.

If we take the eigenvectors $\bv_j$ to have unit norm ($\bv\trasps{j}\bv_j = 1$), then the regression coefficient must satisfy $\bb_j = \bv_j$, and model (\ref{eq:pearpca1}) simplifies to
\begin{equation*}
  \bX = \bX\bv_{1}\bv\trasps{1} + \cdots + \bX\bv_d\bv\trasps{d} + \bE.
\end{equation*}

Hotelling's definition of PCA is the most used one in the literature
and it is the definition of the Least squares solution (\ref{eq:pcaeigenvecs}). Often it is simplified\footnote{This simplification derives from the orthogonality of the eigenvectors of a symmetric matrix and it is equivalent to the orthogonality constraints because, by the definition of eigenvalues, $ 0 = \bp\trasps{j}\bp_k = \bv\trasps{j}\bS\bv_k = \bv\trasps{j}\bv_k \lambda_k$.} by requiring that  $\bv\trasps{j}\bv_j = 1$ and $\bv\trasps{j}\bv_k = 0$ if $j \neq k$ Hence, the loadings vectors are defined as the arguments that maximize
\begin{equation}\label{eq:hotpca}
  \bv\trasps{j}\bS\bv_j = \lambda_j,\, j = 1,\ldots, d;\,\text{subject to  }\bv\trasps{k}\bv_j = \delta_{jk},
\end{equation}
where $\delta_{jk}$ is equal to one if $ j = k$ and to zero otherwise.

Hotelling's definition of PCA does not provide a model on the data or a rationale for which the PCs should be better than other linear combination of the data. As \citet[p 45]{ten} puts it:
\begin{quotation}
``it is undesirable to maximize the variance of the components rather than the variance explained by the components, because only the latter is relevant for the purpose of finding components that summarize the information contained in the variables.''
\end{quotation}
From a practical point of view, it does not matter which of the two definitions of PCA is adopted, because they both give the same solution. However, this is no longer true when sparsity constraints are added to model \ref{eq:pearpca1} because the loadings are not eigenvectors of $\bS$ any more.
\subsection{Sparse PCA}
In this section I will outline how the solutions of SPCA methods are derived but the actual solutions are given in Section \ref{sec:compDet}. Even though in this tutorial I will only consider LS SPCA, I also introduce conventional SPCA to allow readers to appreciate the differences. Most of the results I report are from my \citeyearpar{mer} and \citeyearpar{mer19} papers and I will not always give these references.
\subsubsection{Least Squares SPCA}
The LS SPCA SPCs,  to which I generically refer as LSSPCs, are obtained by adding sparsity constraints to model \ref{eq:pearpca1}. Let $\dX_j$ denote a generic subset of $c_j < p$ variables selected for the $j$--th SPC. Then the SPCs are defined as $\bt_j = \dX_j\da_j$, where $\da_j$ is the vector containing only the nonzero loadings. The standard LS SPCA model with orthogonality constraints, which I call USPCA (U stands for uncorrelated), can be written as
\begin{equation}\label{eq:lspcaunc}
  \bX = \bt_{1}\bb\trasps{1} + \cdots + \bt_d\bb\trasps{d} + \bE,\, \text{subject to } \bp\trasps{j}\bp_k = 0\, \text{if } k > j.
\end{equation}
Just like in ordinary LS regression, the solutions are obtained by minimizing the sum of squared errors $\sum_{i=1}^j\sum_{j = 1}^p e_{ij}^2$. I will refer to the SPCs produced as USPCs.

It is important to notice that the USPCA loadings are neither eigenvalues of $\bS$ nor are orthogonal. For the former reason, the coefficients $\bb_j$ are no longer proportional to the loadings and cannot have unit length. Furthermore, the norms of the SPCs are not equal to the variance that they explain. Consequently, these solutions cannot be simplified as in Hotelling's definition of PCA \ref{eq:hotpca}.

The orthogonality constraints require that the cardinality of each set of sparse loadings is not smaller than its order, which can be undesirable for SPCs of higher order.

Furthermore, such constraints make the computation unstable in some cases. For this reasons, alternative LSSPCs can be obtained by dropping the orthogonality constraints in model (\ref{eq:lspcaunc}); I refer to this model as CSPCA (C stands for correlated) and to the SPCs produced as CSPCs. The CSPCs are computed from Model \ref{eq:lspcaunc} without the orthogonality constraints by sequentially maximizing the \emph{extra} variance explained by each component. Therefore, each CSPC explains the most possible variance of the residuals from the approximations obtained with the preceding CSPCs.
In most cases, the CSPCs of low order are close to the USPCs (the first are ones equal) while the ones of higher order have lower cardinality. The latter may explain slightly more variance, at the price of being correlated with the others. The correlations between CSPCs is generally low and it is inversely related to the proportion of variance that they explain. CSPCs can be computed only for components of higher order, after computing low order orthogonal SPCs.

In \cite{mer19} I propose to compute LSSPCs from the regression of the standard PCs on a subset of the variables. In this approach, called projection SPCA (PSPCA), the PSPCs are obtained by simply solving the regression models
\begin{equation}\label{eq:projspca}
  \bp_j =  \btj g_j + \be_j = \dX_j\da_j + \be_j,\, j = 1,\ldots, d,
\end{equation}
where $g_j$ is a regression coefficient that can be omitted because there no restrictions on the norm of $\da_j$. So, the sparse loadings are simply the coefficients of the regression of $\bX$ on $\dXj$.

If the variables in $\dX_j$ are selected so that the regression coefficient of determination, $R^2$, is  equal to $ \alpha \in(0, 1)$, then the proportion variance explained by the PSPCs with respect to that explained by the PC is not less than $\alpha$.

I call the SPCs obtained by simply regressing the PCs produces \emph{crude} PSPCs. PSPCs that explain more variance and are less correlated can be obtained by regressing the first PC of the orthogonal residuals from the previously computed SPCs (to which I refer simply as PSPCs). In both cases, the PSPCs will be correlated (but the correlation can be decreased by increasing $\alpha$) and will explain less variance than the CSPCs.

Like the CSPCs, the PSPCs of order higher than one may explain more variance than the USPCs, at the price of being correlated with the preceding ones. The computation of the PSPCs is simpler and less computationally expensive than for other LSSPCs.

The main difficulty in computing SPCA is finding good subsets of variables for each SPC, which is well known to be a computationally intractable (NP--hard) problem\footnote{NP stands for nondeterministic polynomial time. It means that an efficient algorithm for solving the problem cannot be found. Because of this, all SPCA methods use \emph{greedy} algorithms that produce suboptimal solutions.}. PSPCA suggests an obvious suboptimal approach to select the variables: use one of the existing variable selection algorithms for regression, which are simple and computationally economical. Once the variables are selected via regression, it is possible to compute USPCs or CSPCs from these. By selecting a minimal $R^2$ threshold, the LSSPCs are guaranteed to explain a proportion not lower than that value of the variance explained by the corresponding PC. In \cite{mer} I suggest also a backward elimination criterion, which gives excellent results but is computationally expensive and tricky to implement. I computed LSSPCs using regression forward selection for very large matrices (as large as 16,000 variables) with computational times below one second per component \citep{mer19}.
\subsubsection{Conventional SPCA}
Conventional SPCA methods are derived by adding sparsity constraints to Hotelling's definition of PCA. Hence, the variance explained by an SPC is measured by its norm, $\bt\trasps{j}\bt_j$, and the loadings are computed by maximizing
\begin{equation}\label{eq:convspca}
  \ba\trasps{j}\bS\ba_j,\, \text{subject to } \bajT\baj = 1\, \text{and card}(\ba_j)\leq k,
\end{equation}
where card($\ba_j$) is the cardinality of $\ba_j$ and $k < p$.
Conventional SPCA methods differ by how they solve this maximization problem. Solutions are obtained numerically under different constraints on the loadings, reviewing which is not necessary for our discussion. A partial review of the plethora of existing methods can be found in \cite{zou18}, for example.
The most popular conventional SPCA methods seem to be that proposed by \cite{zou} with an $L_1$ (Lasso) penalty and its regularized variants. This method is derived from Pearson's PCA Model \ref{eq:lspcaunc} but, since both the coefficient vectors $\daj$ and  $\bb_j$  are constrained to have unit length, the function optimised reduces to the norm of the SPCs (Equation \ref{eq:convspca}) \citep[see][for a proof]{mer}.

The main characteristic of conventional SPCA is that the computed SPCs are simply the PCs of subsets of variables \citep{mog}. To see this, consider that, under sparsity constraints, the SPCs
are equal to $\btj = \dXj\daj$. Therefore, the quantity being maximised in Equation (\ref{eq:convspca}) reduces to
\[
 \ba\trasps{j}\bS\ba_j = \dajT\dXjT\dXj\daj = \dajT\bdS_{j}\daj,
\]
where $\dSj \propto \dXjT\dXj$. Hence, the solutions to the maximization \ref{eq:convspca} is the loading vector of the first PC of $\dXj$ augmented with zeroes for the missing variables.
The maximization of this objective function requires selecting variables that are as highly correlated as possible. Furthermore, the optimization concerns only the selected subset of variables while the rest of the variables is ignored.

The maximization of the norm of the SPCs leads to erroneous results. For example, linear combinations of perfectly correlated variables are considered to explain more variance than linear combinations of fewer of them. As an example, assume that we observed five perfectly correlate variables $\bx_j$, each with variance equal to $j = 1,\ldots, 5$. Since the data and covariance matrices have rank equal to one, the first PC is enough to explain all the variance of the data. This PC is proportional to any one of the variables, and to any linear combination of them, so also to any SPC. In spite of this, the norm of conventional SPCs increases with their cardinality, as shown in Figure \ref{fig:ex_load_scores}. An even more bizarre example can be obtained by standardizing these variables to the same norm (hence they become identical), as shown in \cite{mer19}. This behavior is observed, with due differences, also with less than perfectly correlated variables.
\begingroup
\renewcommand{\arraystretch}{0.7}
\begin{figure}[H]
  \centering
\begin{subtable}[h]{0.5\textwidth}
 \centering
\scriptsize{
\begin{tabular}{lrrrrrr}
\toprule
     & \multicolumn{4}{c}{SPCs}& &\multicolumn{1}{c}{PC} \\
     \cmidrule{2-5}\cmidrule{7-7}
      & \multicolumn{6}{c}{cardinality} \\
\cmidrule{2-7}
 variable     & 1     & 2     & 3     & 4  &   & 5 \\
\midrule
$x_1$ & 0     & 0     & 0     & 0     &       & 0.26 \\
$x_2$ & 0     & 0     & 0     & 0.38  &       & 0.37 \\
$x_3$ & 0     & 0     & 0.5   & 0.46  &       & 0.45 \\
$x_4$ & 0     & 0.67  & 0.58  & 0.53  &       & 0.52 \\
$x_5$ & 1     & 0.75  & 0.65  & 0.6   &       & 0.58 \\
\midrule
norm  & 5  & 9   & 12  & 14  && 15 \\
rel. norm & 0.33  & 0.60  & 0.80  & 0.93  && 1.0 \\
\bottomrule
\end{tabular}%
}
\label{tab:expl}%
\end{subtable}\hfill%
\begin{subfigure}[h]{0.5\textwidth}
\includegraphics[width = 0.66\textwidth]{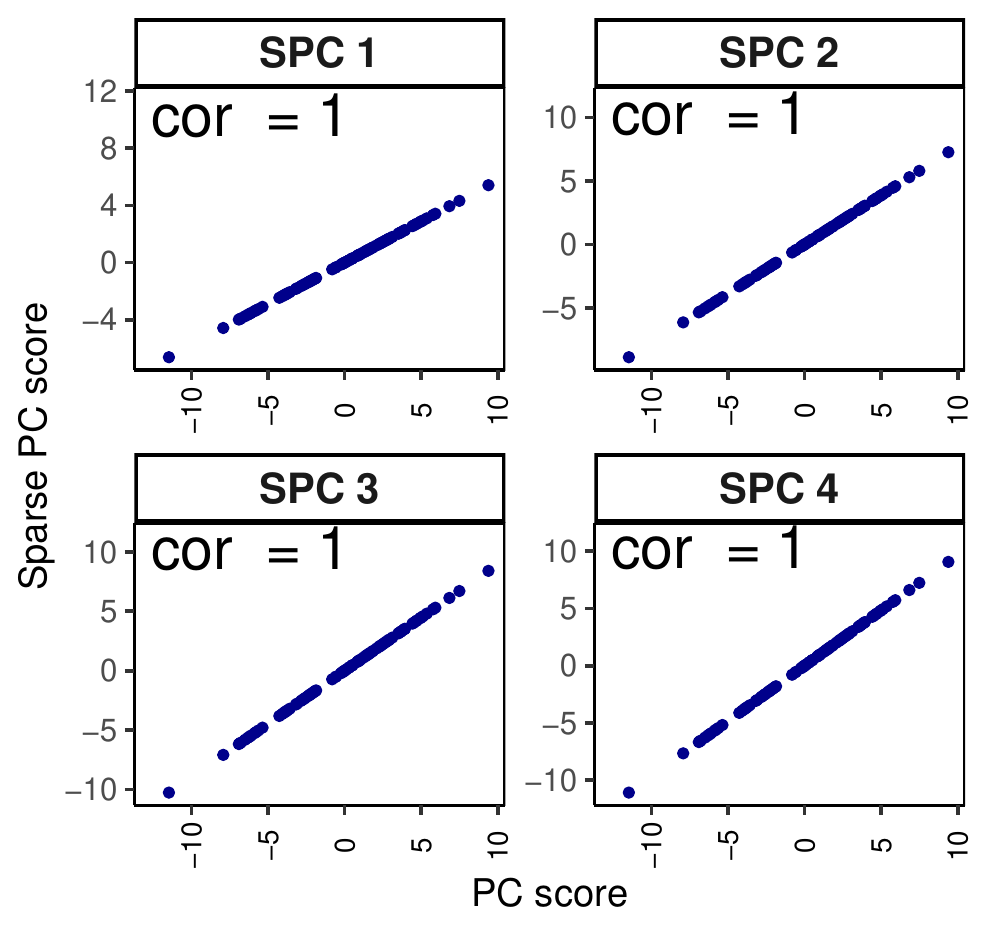}
        \label{fig:sub2}
    \end{subfigure}
    \caption{Table with the loadings and norms, and the scatter plots of the scores against the first PC of conventional SPCs of different cardinalities for a set of five perfectly correlated variables.}
    \label{fig:ex_load_scores}
\end{figure}
\endgroup
Another drawback of conventional SPCA methods regards sets of linearly dependent variables (but not necessarily pairwise linearly dependent). This is the case, for example, when there are fewer observations than variables and the rank of the data matrix is equal to the number of observations\footnote{The rank is actually equal to the number of observations minus one when the variables are centered to zero mean.}.
When the rank of the data matrix, $r$, is lower than the number of variables, any linear combination of the variables (including the PCs) can be expressed as a linear combination of a subset of $r$ linearly independent variables. Nonetheless, in the literature there are several examples of SPCs with cardinality much larger than the matrix rank. For example, the applications of SPCA on the 16,063 genes and 144 samples ``Ramaswamy'' data in \citet{zou, wan}, among others, where SPCs with cardinality in the thousands are computed. These can be compared with the results of LS SPCA on the same dataset in \citet{mer19}, where an SPC with cardinality equal to 143 perfectly reproduces the first PC. All this discussion should be a convincing proof that conventional SPCA methods are not the best choice for sparsifying the PCs.
%
\section{What to expect from LS SPCA}
In LS SPCA the variables forming the sparse components are selected only to maximize the variance explained. Consequently, LS SPCA can be very useful to identify a few key variables able to summarize the whole set when combined together. However, there is no guarantee that the SPCs will be easy to ``interpret'' or that are combinations of variables measuring similar quantities. This is due to the fact that the  variables selected tend to have low multiple correlation, because in a least squares framework correlation is equivalent to redundancy. Instead, variables forming ``valid constructs'' are required to have high multiple correlations.

However, interpretability is a subjective concept. A case in point
are the first thresholded PC and USPC computed on baseball hitters playing statistics\footnote{See details about this  dataset in Section \ref{sec:bsbl}}. The corresponding percentage contributions are equal to

\begingroup
\renewcommand\baselinestretch{0.7}
\noindent
{\small
\textbf{thresholded PC}\\
    12.5\%\text{(years in major leagues)} + 14.6\%\text{(times at bat in career)} + 14.6\%\text{(hits in career)} +\\
    14.1\%\text{(home runs in career)} +
    15\%\text{(runs in career)} +
    15.1\%\text{(runs batted-in in career)} +\\
    14.1\%\text{(walks in career)}\\
  \textbf{USPC}\\
   39.9\%\text{(runs batted-in in 1986)} + 91.7\%\text{(runs in career)}
   }
\endgroup

Some analysts may consider the thresholded PC to be easier to interpret because it summarises a player's career performance. Others may consider the USPC to be more meaningful because it gives a comprehensive summary of the performance of a player using only two key playing statistics.

However, objectively, the LSSPC has better properties than the thresholded PC. In fact, the thresholded PC explains  $93.4\% $ of the variance explained by the first PC and is a combination of seven (all available) career statistics, which have multiple correlations equal to $0.86, 0.92, 0.97, 0.99, 0.99, 0.99$ and $1.00$. In contrast, the LSSPC explains 97.4\% of the variance explained by the first PC and is a combination of just two variables which have correlation equal to $0.31$.

The LS SPCA solutions are not necessarily globally optimal.
Globally optimal solutions for SPCA are computationally too demanding to be found because for $d$ components it would be necessary to evaluate $(2^p -1)^d$ solutions. Therefore, locally optimal solutions are computed by selecting the variables sequentially for each component. When the number of variables is large, also the local solutions must necessary be found with suboptimal algorithms. This is, for example, the case when using regression variable selection algorithms when the \emph{all--subsets} exhaustive search becomes too computationally demanding and greedy selection algorithms are required.

One last consideration regards the variance explained by a set of SPCs. The USPCs explain the most possible variance under orthogonality constraints. However, it is possible to find sets of correlated SPCs, with same or lower cardinality, which explain more net variance than the USPCs. Therefore, if orthogonality (or low correlation) is not important for the analysis, other methods can be used in the hope of finding sets of correlated SPCs that explain the variance more parsimoniously.

\section{Demonstrative examples}\label{sec:example}
In this section I give a few examples and comparisons of LS SPCA applied on two data sets.
I chose these two datsets because they have different correlation structures. The Baseball Hitters dataset has a clear correlation structure and is easy to analyze with PCA. Instead, the Students Ability data has a week correlation structure but three of the variables have a much higher variance than the others. So, it is difficult to analyze with PCA. I will not attempt to give interpretations because I am not an expert in baseball or Psychometry and my aim is simply to illustrate the results of LS SPCA.

For reporting the results I show loadings scaled to percentage contributions (sum of the absolute values equal to one). The subsets of variables for LSSPCs are selected as the smallest subset giving $R^2 \geq \alpha$ in in regressing the PC (with an exhaustive search, unless differently specified). So, I will simply refer to $\alpha$ to characterize the SPCs. As a measure of goodness of fit for the SPCs I use the proportion of cumulative variance explained by a set of SPCs with respect to that explained by the corresponding standard PCs. This is denoted by \emph{RCVEXP}.
\subsection{Baseball Hitters data}\label{sec:bsbl}
\begin{figure}[H]
\centering
\includegraphics[width = 0.45\textwidth]
 {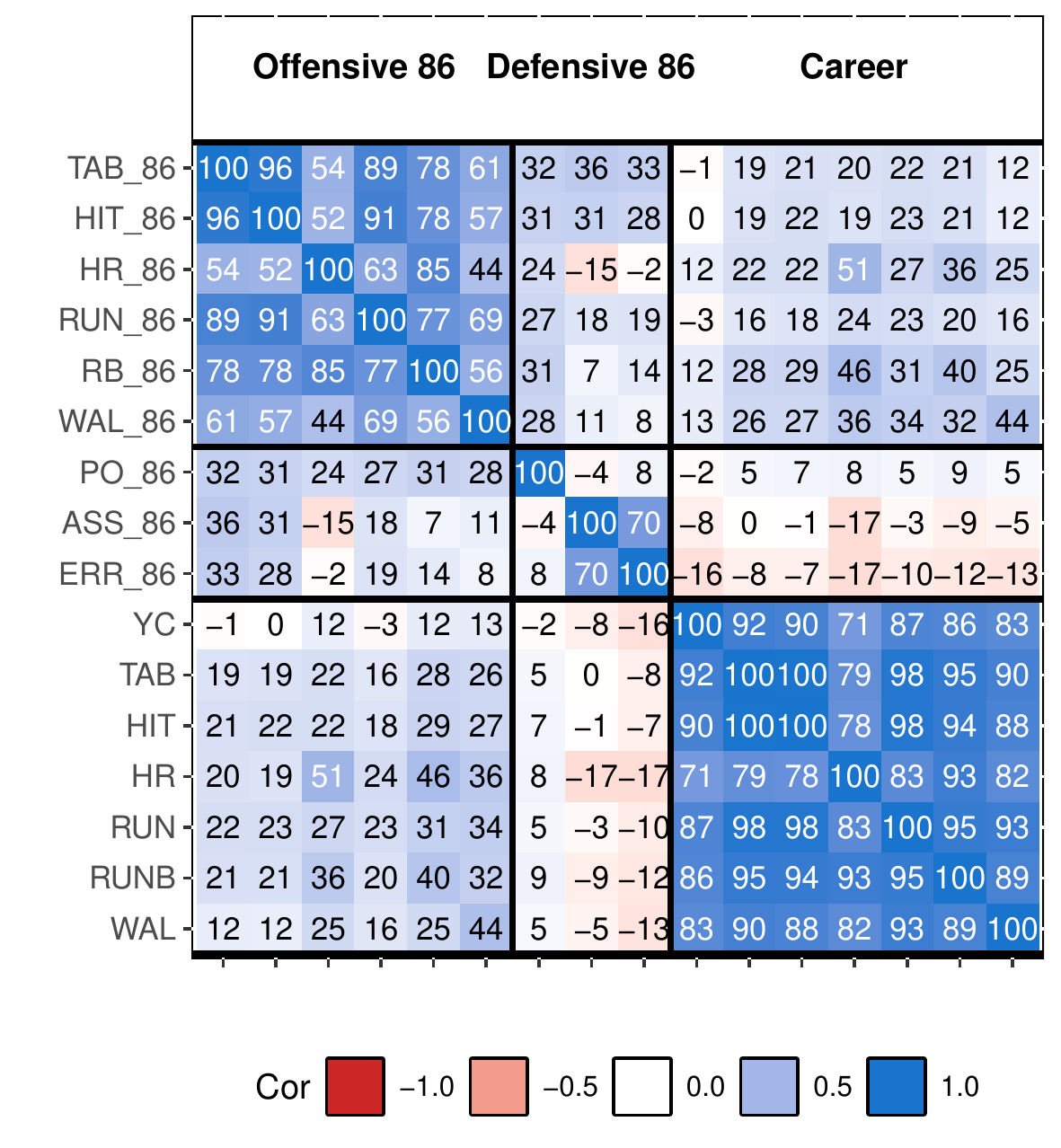}
\caption{Correlation between the hitters data variables expressed as percentage.}%
\label{fig:hitters_cor}
\end{figure}

This dataset contains 16 playing statistics relative to 263 Major League baseball players (hitters), nine  recorded in 1986 and seven over their whole career. The variables are indicators of the players' offensive play in 1986 and during their career (six and seven, respectively), the remaining three are indicators of the players' defensive play in 1986, as shown in Table \ref{tab:hitvars}. Playing statistics of different type are correlated among themselves and much less with the others, as shown in Figure \ref{fig:hitters_cor}.  Since the playing statistics are nonhomogeneous measures, I ran the analyses on the variables scaled to unit variance. Therefore, the loadings are computed from the correlation matrix.
\begingroup
\renewcommand{\arraystretch}{0.75}
\begin{table}[H]
  \centering
  \caption{Variables in the Hitters dataset}
  {\scriptsize
\begin{tabular}{llrrrrll}
\multicolumn{2}{l}{Offensive play in 1986 (OFF 86)} &       & \multicolumn{2}{l}{Defensive play in 1986} (DEF 86)&       & \multicolumn{2}{l}{Offensive play in career} (OFF CAR)\\
\cmidrule{1-2}\cmidrule{4-5}\cmidrule{7-8}Label & Name  &       & \multicolumn{1}{l}{Label} & \multicolumn{1}{l}{Name} &       & Label & Name \\
\cmidrule{1-2}\cmidrule{4-5}\cmidrule{7-8}      &       &       &       &       &       & YC    & years in the major leagues \\
TAB\_86 & times at bat in 1986 &       & \multicolumn{1}{l}{PO\_86} & \multicolumn{1}{l}{put outs in 1986} &       & TAB   & times at bat during his career \\
HIT\_86 & hits in 1986 &       & \multicolumn{1}{l}{ASS\_86} & \multicolumn{1}{l}{assists in 1986} &       & HIT   & hits during his career \\
HR\_86 & home runs in 1986 &       & \multicolumn{1}{l}{ERR\_86} & \multicolumn{1}{l}{errors in 1986} &       & HR    & home runs during his career \\
RUN\_86 & runs in 1986 &       &       &       &       & RUN   & runs during his career \\
RB\_86 & runs batted-in in 1986 &       &       &       &       & RUNB  & runs batted-in during his career \\
WAL\_86 & walks in 1986 &       &       &       &       & WAL   & walks during his career \\
\bottomrule
\end{tabular}%
}
\label{tab:hitvars}%
\end{table}
\endgroup

\subsubsection{PCA}
\begin{figure}[H]
\caption{Contributions of the first four PCs of the Hitters data}%
\label{fig:bsbl_4pcloads}
\centering
\includegraphics[width = 1\textwidth]
 {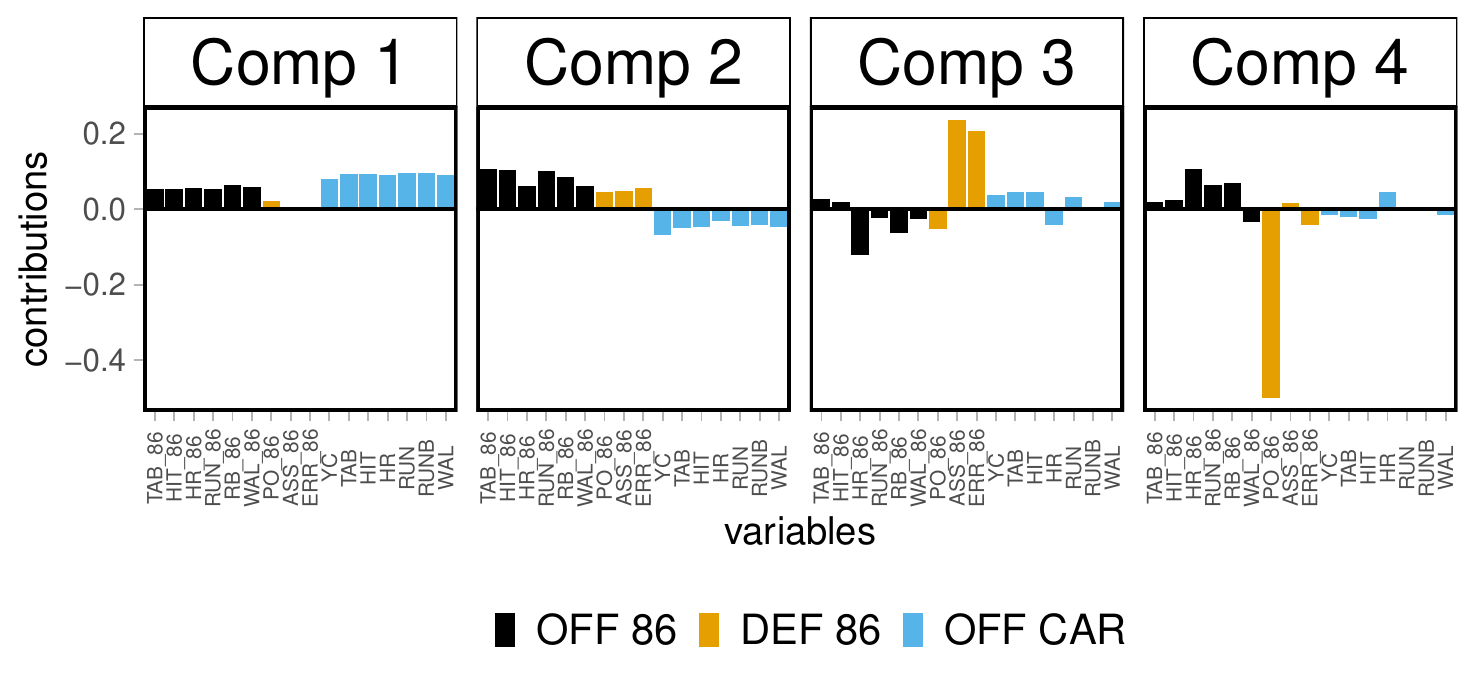}
\end{figure}

Figure \ref{fig:bsbl_4pcloads} shows the contributions of the first four PCs of this dataset.
Even though for the first two PCs, the statistics of the same type have loadings of the same sign, it would be difficult to describe these combinations of variables in detail with one sentence.
\subsubsection{LS SPCA}
Table \ref{fig:bsbl_uspca3alphas} shows the summaries comparing the first four orthogonal USPCs computed with decreasing  values of $\alpha$ equal to 0.99, 0.95 and 0.90 and selecting the variables by exhaustive search. The 99\% USPCs have higher cardinality than the others, the 90\% USPCs have lower cardinality than the 95\% ones only in the second component. All 90\% components have rCvexp index higher than 95\%, showing that 90\% rCvexp cannot be reached with lower cardinality.

\begin{figure}[H]
\centering
\includegraphics[width = 1\textwidth]
 {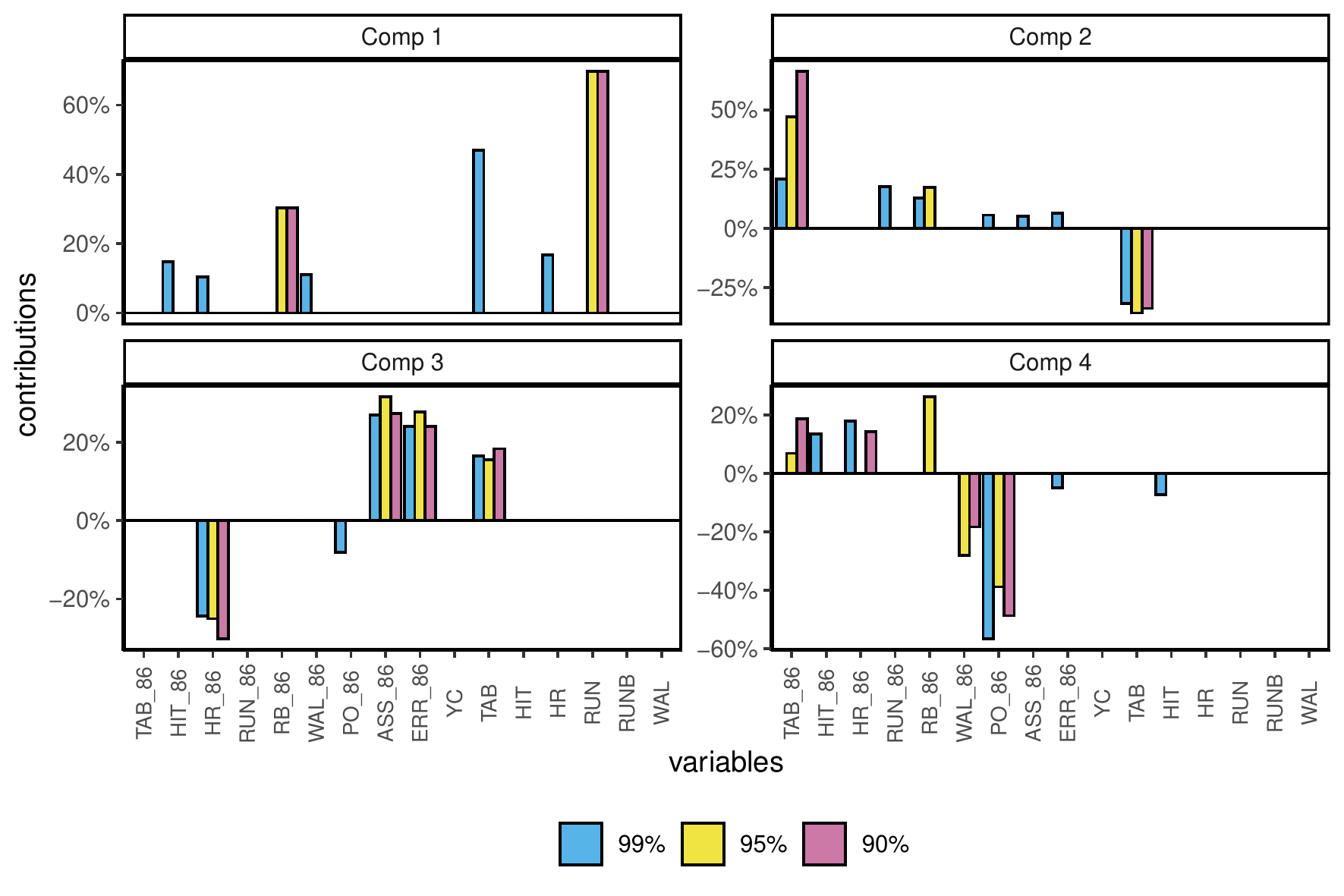}
\caption{Contributions of the first four USPCs computed requiring RCVEXP $> 99\%,\, 95\%$ and $90\%$ respectively for the baseball hitters data.}%
\label{fig:bsbl_uspcaE_diffAlpha}
\end{figure}

\begingroup
\renewcommand{\arraystretch}{0.6}
\begin{table}[H]
  \centering
\caption{Contributions of the first four USPCs computed requiring RCVEXP $> 99\%,\, 95\%$ and $90\%$, respectively, and selecting the variables with exhaustive search for the baseball hitters data.}%
\label{fig:bsbl_uspca3alphas}  {\scriptsize
    \begin{tabular}{lrrrrrrrrrrrrrrr}
    \toprule
          & \multicolumn{3}{c}{1st Component} &       & \multicolumn{3}{c}{2nd Component} &       & \multicolumn{3}{c}{3rd Component} &       & \multicolumn{3}{c}{4th Component} \\
\cmidrule{2-4}\cmidrule{6-8}\cmidrule{10-12}\cmidrule{14-16}    $\alpha$ & 99\%  & 95\%  & 90\%  &       & 99\%  & 95\%  & 90\%  &       & 99\%  & 95\%  & 90\%  &       & 99\%  & 95\%  & 90\% \\
    \midrule
    VEXP  & 44.9  & 44    & 44    &       & 25.5  & 24.7  & 24.1  &       & 10.7  & 10.8  & 10.6  &       & 5.4   & 5.5   & 5.6 \\
    CVEXP & 44.9  & 44    & 44    &       & 70.4  & 68.7  & 68.1  &       & 81.1  & 79.5  & 78.7  &       & 86.6  & 85    & 84.3 \\
    RCVEXP & 99.5  & 97.4  & 97.4  &       & 99.4  & 96.9  & 96.1  &       & 99.4  & 97.4  & 96.4  &       & 99.4  & 97.6  & 96.9 \\
    Card  & 5     & 2     & 2     &       & 7     & 3     & 2     &       & 5     & 4     & 4     &       & 5     & 4     & 4 \\
    \bottomrule
    \end{tabular}%
}
\label{tab:bsbl_u3alphas}%
\end{table}%
\endgroup
The loadings of three sets of USPCs are plotted in Figure \ref{fig:bsbl_uspcaE_diffAlpha}.
The variables selected for the first three 90\% USPCs are the same or subsets of those selected for the 95\% USPCs, while for the fourth one of the variables is different. Only for the second and third 99\% USPCs
the sets of variables selected contain the variables selected for the other SPCs.
%
%
Table \ref{tab:bsbl_c3alphas} shows the summaries comparing the first four correlated CSPCs computed with decreasing  values of $\alpha$ equal to 0.99, 0.95 and 0.90 and selecting the variables by forward selection.
The first three sets of SPCs are almost identical.
The only substantial difference is in the fourth set, where the USPC is a combination of four variables (as required by orthogonality), whereas the CSPC has cardinality two.

\begingroup
\renewcommand{\arraystretch}{0.6}
\begin{table}[H]
  \centering
\caption{Contributions of the first four CSPCs computed requiring RCVEXP $> 99\%,\, 95\%$ and $90\%$, respectively, and selecting the variables with forward selection for the baseball hitters data.}%
 {\scriptsize
    \begin{tabular}{lrrrrrrrrrrrrrrr}
    \toprule
          & \multicolumn{3}{c}{1st Component} &       & \multicolumn{3}{c}{2nd Component} &       & \multicolumn{3}{c}{3rd Component} &       & \multicolumn{3}{c}{4th Component} \\
\cmidrule{2-4}\cmidrule{6-8}\cmidrule{10-12}\cmidrule{14-16}    $\alpha$ & 99\%  & 95\%  & 90\%  &       & 99\%  & 95\%  & 90\%  &       & 99\%  & 95\%  & 90\%  &       & 99\%  & 95\%  & 90\% \\
    \midrule
    VEXP  & 44.8  & 43.9  & 43.9  &       & 25.5  & 24.7  & 24.2  &       & 10.7  & 10.8  & 10.6  &       & 5.5   & 5.5   & 5.5 \\
    CVEXP & 44.8  & 43.9  & 43.9  &       & 70.3  & 68.5  & 68    &       & 81.1  & 79.4  & 78.7  &       & 86.5  & 84.9  & 84.2 \\
    RCVEXP & 99.4  & 97.2  & 97.2  &       & 99.3  & 96.8  & 96    &       & 99.3  & 97.2  & 96.4  &       & 99.4  & 97.5  & 96.7 \\
    Card  & 5     & 2     & 2     &       & 7     & 3     & 2     &       & 5     & 4     & 4     &       & 5     & 2     & 2 \\
    \bottomrule
    \end{tabular}%
}
\label{tab:bsbl_c3alphas}%
\end{table}%
\endgroup
The correlation between the CSPCs is negligible with a maximum equal to 0.12 between the second and fourth 90\% CSPCs. This shows that relaxing the orthogonality requirements may produce to more efficient solutions. In some cases the correlations between CSPCs are considerable. Increasing $\alpha$ reduces these correlations.
%
%

Table \ref{tab:bsbl_loaduc3alphas} shows the contributions of the first two USPCs together with those of the corresponding CSPCs. Since the CSPCs were computed by selecting the variables with forward selections, the variables selected for SPCs with lower $\alpha$ are subsets of those selected for SPCs with larger $\alpha$. This is not always the case when the variables are selected with exhaustive search, as I did for the USPCs.
\begingroup
\renewcommand{\arraystretch}{0.6}
\begin{table}[H]
  \centering
  \caption{Contributions of the first two USPCs and CSPCs computed using with exhaustive search and forward selection, respectively and requiring RCVEXP $> 99\%,\, 95\%$ and $90\%$ for the baseball hitters data.}
  {\scriptsize
\begin{tabular}{lrrrrrrrrrrrrrrr}
\toprule
      & \multicolumn{7}{c}{1st Component}                     &       & \multicolumn{7}{c}{2nd Component} \\
\cmidrule{2-8}\cmidrule{10-16}      & \multicolumn{3}{c}{USPCA exhaustive} &       & \multicolumn{3}{c}{CSPCA forward} &       & \multicolumn{3}{c}{USPCA exhaustive} &       & \multicolumn{3}{c}{CSPCA forward} \\
\cmidrule{2-4}\cmidrule{6-8}\cmidrule{10-12}\cmidrule{14-16}$\alpha$ & 99\%  & 95\%  & 90\%  &       & 99\%  & 95\%  & 90\%  &       & 99\%  & 95\%  & 90\%  &       & 99\%  & 95\%  & 90\% \\
\midrule
TAB\_86 &       &       &       &       &       &       &       &       & 20.8  & 46.9  & 66.2  &       & 20.9  & 46.2  & 63.2 \\
HIT\_86 & 14.7  &       &       &       &       &       &       &       &       &       &       &       &       &       &  \\
HR\_86 & 10.4  &       &       &       &       &       &       &       &       &       &       &       &       &       &  \\
RUN\_86 &       &       &       &       & 16.9  & 28.1  & 28.1  &       & 17.7  &       &       &       & 17.9  &       &  \\
RB\_86 &       & 30.3  & 30.3  &       & 15.2  &       &       &       & 12.8  & 17.3  &       &       & 13    & 16.7  &  \\
WAL\_86 & 11    &       &       &       &       &       &       &       &       &       &       &       &       &       &  \\
PO\_86 &       &       &       &       &       &       &       &       & 5.5   &       &       &       & 5.6   &       &  \\
ASS\_86 &       &       &       &       &       &       &       &       & 5.1   &       &       &       & 5.1   &       &  \\
ERR\_86 &       &       &       &       &       &       &       &       & 6.3   &       &       &       & 6.3   &       &  \\
YC    &       &       &       &       &       &       &       &       &       &       &       &       &       &       &  \\
TAB   & 47    &       &       &       & 24.9  &       &       &       & -31.8 & -35.7 & -33.8 &       & -31.2 & -37.1 & -36.8 \\
HIT   &       &       &       &       &       &       &       &       &       &       &       &       &       &       &  \\
HR    & 16.8  &       &       &       &       &       &       &       &       &       &       &       &       &       &  \\
RUN   &       & 69.7  & 69.7  &       &       &       &       &       &       &       &       &       &       &       &  \\
RUNB  &       &       &       &       & 25.3  & 71.9  & 71.9  &       &       &       &       &       &       &       &  \\
WAL   &       &       &       &       & 17.7  &       &       &       &       &       &       &       &       &       &  \\
\midrule
Card  & 5     & 2     & 2     &       & 5     & 2     & 2     &       & 7     & 3     & 2     &       & 7     & 3     & 2 \\
CVEXP & 44.9  & 44    & 44    &       & 44.8  & 43.9  & 43.9  &       & 70.4  & 68.7  & 68.1  &       & 70.3  & 68.5  & 68 \\
\bottomrule
\end{tabular}%
}
\label{tab:bsbl_loaduc3alphas}%
\end{table}%
\endgroup
%

\subsection{comparison with thresholding}
\begin{figure}[H]
\centering
\includegraphics[width = 0.66\textwidth]
 {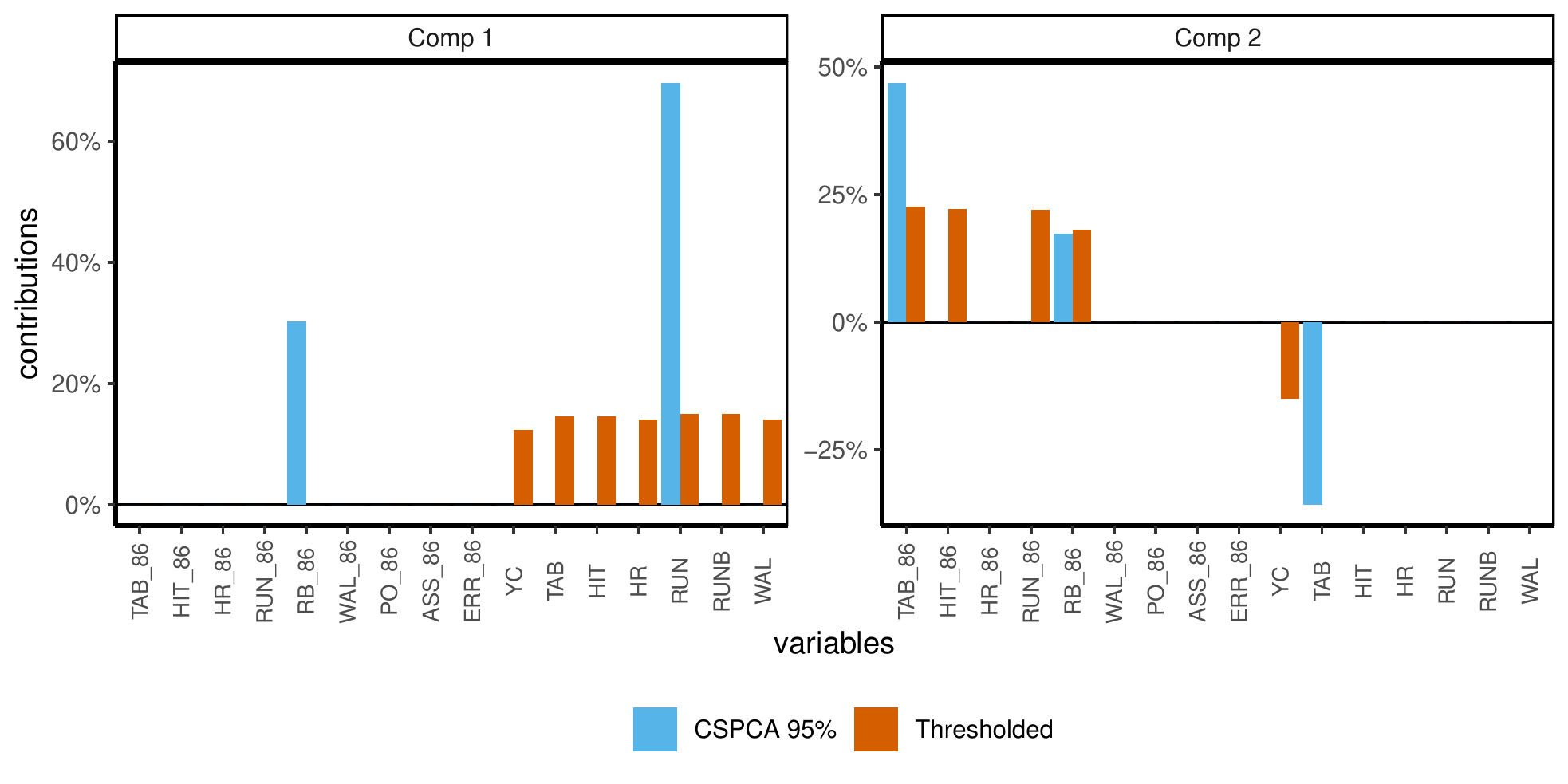}
\caption{Contributions of the first two tresholded PCs and USPCs for the baseball hitters data.}%
\label{fig:bsbl_compthresh}
\end{figure}

\begin{figure}[H]
\centering
\includegraphics[width = 0.66\textwidth]
 {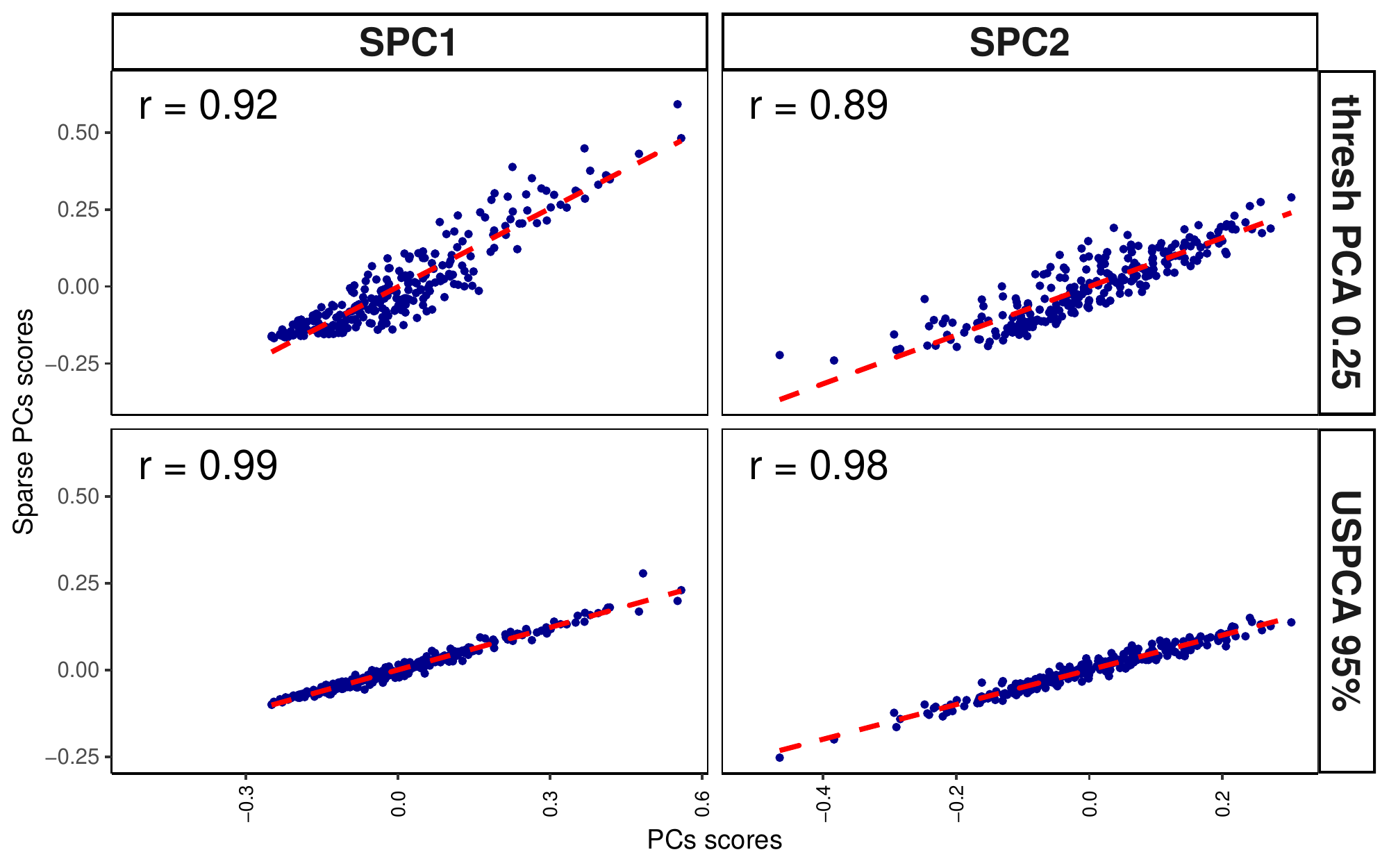}
\caption{Scatter plot of the first two tresholded PCs (top) and LS SPCA scores (bottom) against the corresponding PCs' scores.}%
\label{fig:bsbl_scores}
\end{figure}

LS SPCA produces a much closer approximation to the PCs than thresholding. The contributions of the of th efirst two PCs thresholded at 0.25 and the USPCs 95\% are shown in Figure \ref{fig:bsbl_compthresh}.
Figure \ref{fig:bsbl_scores} shows the scatter plots of the scores of the first two PCs thresholded with threshold $0.25$ and those of the first two 95\% USPCs  against the scores of the corresponding PCs. The USPCs have a much higher correlation with the corresponding PCs than the thresholded PCs with lower cardinality, as shown in Table \ref{tab:bsblu+thrstatsl}.

\begingroup
\renewcommand{\arraystretch}{0.6}
\begin{table}[H]
  \centering
  \caption{Summary statistics of the first two thresholded and USPCA 95\% SPCs computed on the Baseball Hitters data.}
  {\scriptsize
    \begin{tabular}{lrrrrr}
    \toprule
          & \multicolumn{2}{c}{1st Component} &       & \multicolumn{2}{c}{2nd Component} \\
\cmidrule{2-3}\cmidrule{5-6}          & \multicolumn{1}{l}{thresh PCA } & \multicolumn{1}{l}{USPCA 95\%} &       & \multicolumn{1}{l}{thresh PCA } & \multicolumn{1}{l}{USPCA 95\%} \\
\cmidrule{2-6}
VEXP  & 42.2 & 43.9 &       & 27.6 & 24.7 \\
    CVEXP & 42.2 & 43.9 &       & 69.7 & 68.5 \\
    RCVEXP & 93.4 & 97.2 &       & 98.5 & 96.8 \\
    Card  & 7     & 2     &       & 5     & 3 \\
    MinCont & 12.5 & 28.1 &       & 15.0 & 16.8 \\
    \bottomrule
    \end{tabular}%
}
  \label{tab:bsblu+thrstatsl}%
\end{table}%
\endgroup

However, most of the variables selected by thresholding present high pairwise correlation, and even more importantly, extremely high multiple correlation, as shown in Table \ref{tabthrpca-vifs}. This means that some of these variables are redundant and do not contribute to explaining the variance of the data.
\begingroup
\renewcommand{\arraystretch}{0.6}
\begin{table}[H]
  \centering
  \caption{Multiple correlation coefficients among variables selected by thresholding for the first two PCs.}
{\scriptsize
\begin{tabular}{rrrrrrr}
\multicolumn{3}{l}{First component} &       &       &       &  \\
\midrule
\multicolumn{1}{l}{YC} & \multicolumn{1}{l}{WAL} & \multicolumn{1}{l}{HR} & \multicolumn{1}{l}{RUN} & \multicolumn{1}{l}{RUNB} & \multicolumn{1}{l}{TAB} & \multicolumn{1}{l}{HIT} \\
0.86  & 0.92  & 0.97  & 0.99  & 0.99  & 0.99  & 1 \\
\midrule
\multicolumn{3}{l}{Second component} &       &       &       &  \\
\midrule
\multicolumn{1}{l}{YC} & \multicolumn{1}{l}{RB\_86} & \multicolumn{1}{l}{RUN\_86} & \multicolumn{1}{l}{TAB\_86} & \multicolumn{1}{l}{HIT\_86} &       &  \\
0.06  & 0.66  & 0.84  & 0.93  & 0.94  &       &  \\
\bottomrule
\end{tabular}%
}
  \label{tabthrpca-vifs}%
\end{table}%
\endgroup

\subsection{Students Ability data}\label{sec:ferdata}
\begin{figure}[H]
\centering
\includegraphics[width = 0.9\textwidth]
 {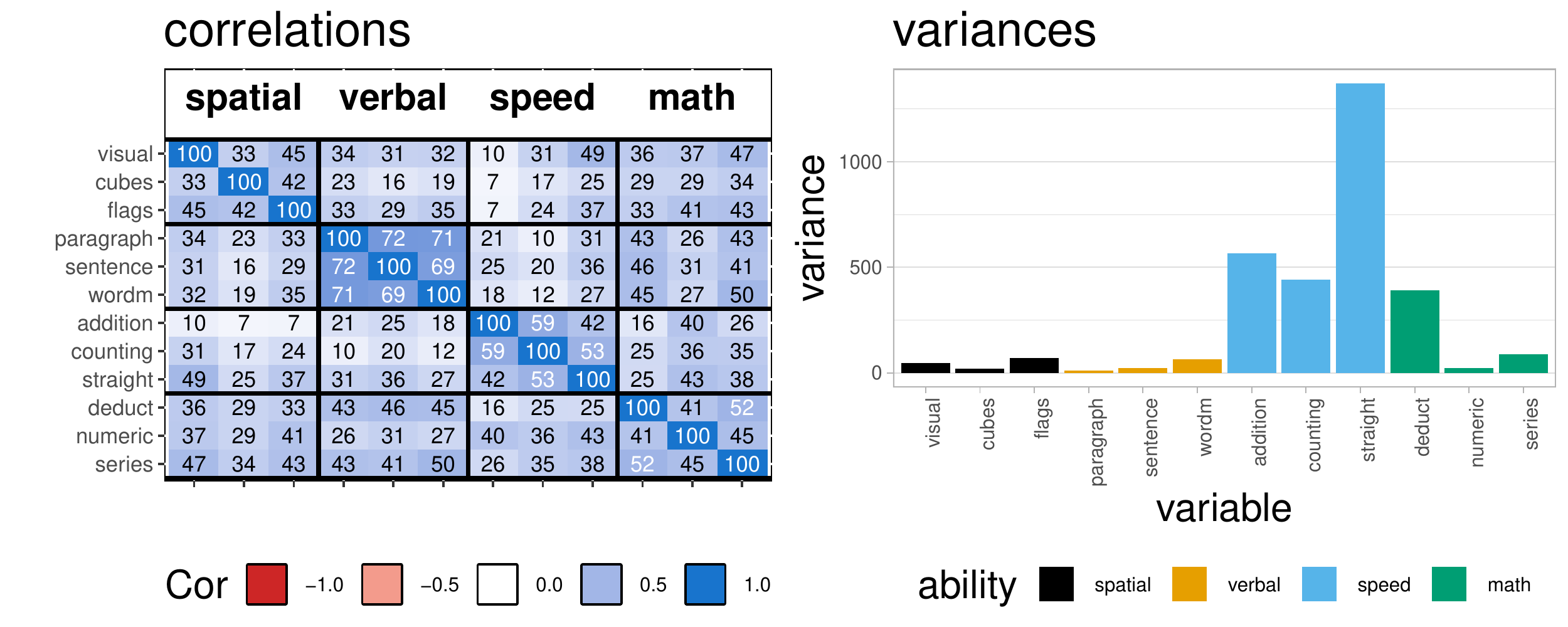}
\caption{Correlation between the Students Ability dataset variables expressed as percentages (left) and the  variances of the variables (right).}%
\label{fig:fer_corvar}
\end{figure}

This classic dataset contains the results of ability tests taken by grade six and seven students. It was first described in \cite{hol} and it has been analyzed in several subsequent papers\footnote{For a partial review of some applications see \href{https://www.rdocumentation.org/packages/psychTools/versions/2.0.8/topics/holzinger.swineford}{the psychTools package documentation}.}. I use the same subset of 12 tests used in \cite{fer}. Details can be found in the papers just mentioned. The tests considered and labels that I use are shown in Table \ref{tab:ferdatadesc}.
\begingroup\renewcommand{\arraystretch}{0.6}
\begin{table}[H]
  \centering
  \caption{Tests considered in the Students Ability dataset.}
  {\scriptsize
    \begin{tabular}{rllll}
    \toprule
    \multicolumn{1}{l}{No.} & Test name & Ability Label & Ability & Test description \\
    \midrule
    1     & visual & SPL   & spatial & Visual perception test \\
    2     & cubes & SPL   & spatial & Cubes simplification \\
    3     & flags & SPL   & spatial & Flags visual discrimination test \\
    4     & paragraph & VBL   & verbal & Paragraph comprehension test \\
    5     & sentence & VBL   & verbal & Sentence completion test \\
    6     & wordm & VBL   & verbal & Word meaning test \\
    7     & addition & SPD   & speed & Addition test \\
    8     & counting & SPD   & speed & Counting of dots in a shape \\
    9     & straight & SPD   & speed & Discriminating straight and curved lines \\
    10    & deduct & MTH   & mathematical & Deduction test \\
    11    & numeric & MTH   & mathematical & Numeric test \\
    12    & series & MTH   & mathematical & Numerical series test \\
    \bottomrule
    \end{tabular}%
  }
  \label{tab:ferdatadesc}%
\end{table}%
\endgroup
This battery of tests shows low internal validity because the test scores in each ability (with the exception of verbal) are weakly correlated among each other and have similar correlation with tests of other abilities, as shown in Figure \ref{fig:fer_corvar}. Moreover, the scores of the three speed tests and the \emph{deduction} test have a much larger variance than the other scores, as shown in Figure \ref{fig:fer_corvar}. The pooled variance of these four variables alone accounts for about 90\% of the total variance of the dataset.
Since the test scores are on the same scale, PCA should be run on the unscaled variables. The first PCs
and 95\% LSSPC computed on the covariance matrix well approximate the data because of the presence of the four variables with dominating variance. The contributions of the resulting first USPC are shown in Figure \ref{fig:intro_loads}.

Next, I will apply LS SPCA to the variables scaled to unit variance to illustrate the behaviour of LS SPCA on a set of weakly correlated variables, which is difficult to well approximate with few PCs. Other authors analyzed the dataset applying PCA on the scaled variables.

\begin{figure}[H]
\caption{Contributions of the first four PCs of the Students Ability dataset}%
\label{fig:fers_PCloads}
\centering
\includegraphics[width = 1\textwidth]
 {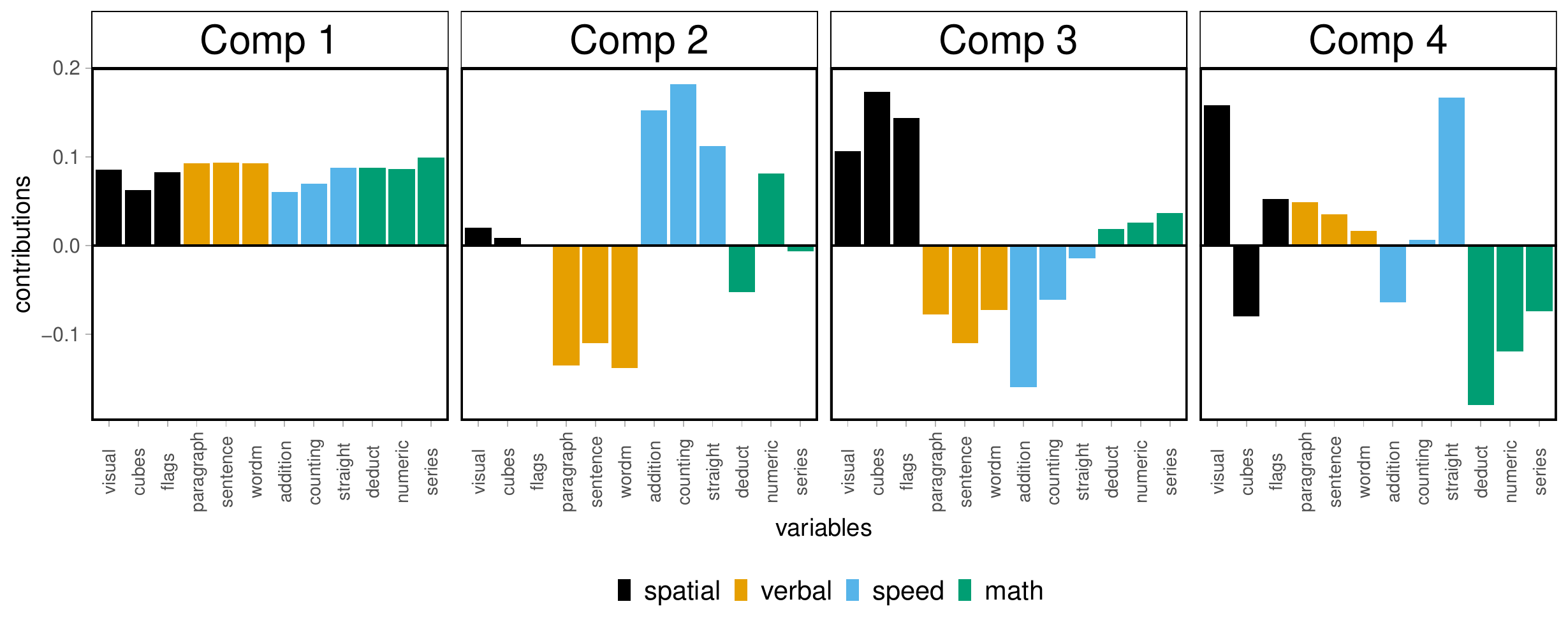}
\end{figure}

\begin{figure}[H]
\centering
\includegraphics[width = 0.75\textwidth]
 {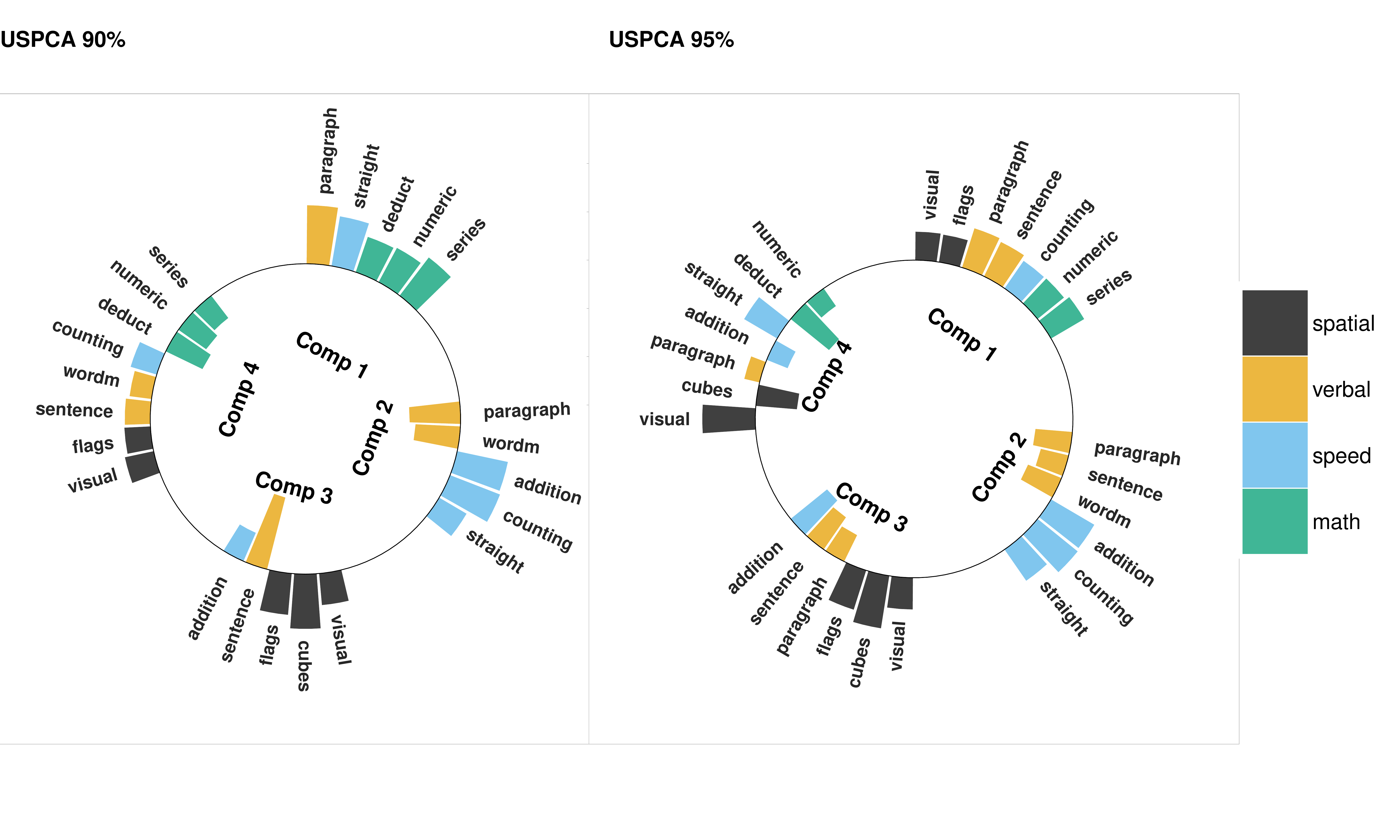}
\caption{Contributions of the first four 90\% and 95\% USPCs for the Students Ability dataset.}%
\label{fig:fer_u9095circ}
\end{figure}

Figure \ref{fig:fers_PCloads} shows the contributions of the first four PCs. The first PC is roughly the average of all variables and the second is mainly the differences between verbal and speed abilities with contradictory contributions from two of the mathematical ability tests.
Figure \ref{fig:fer_u9095circ} shows the contributions of the 90\% and 95\% USPCs. As expected, the USPCs are not very parsimonious and the nonzero loadings correspond to variables in different ability types. However, there is a noticeable simplification in comparison with the PCs.

The summary statistics shown in Table \ref{tab:fers_loads} indicate that the marginal increase in variance explained by the 95\% USPCs is small compared to the increase in cardinality.
The USPCs in both sets are highly correlated with the corresponding PCs, as shown in Table \ref{tab:fer_corpcs}.
\begingroup\renewcommand{\arraystretch}{0.6}
\begin{table}[H]
  \centering
  \caption{Summary statistics for the first four 90\% and 95\% USPCs for the Students Ability dataset.}
{\scriptsize
    \begin{tabular}{lrrrrrrrrrrr}
    \toprule
          & \multicolumn{2}{c}{\textbf{Comp 1}} &       & \multicolumn{2}{c}{\textbf{Comp 2}} &       & \multicolumn{2}{c}{\textbf{Comp 3}} &       & \multicolumn{2}{c}{\textbf{Comp 4}} \\
\cmidrule{2-12}
$\alpha$  & \multicolumn{1}{l}{ 90\%} & \multicolumn{1}{l}{ 95\%} &       & \multicolumn{1}{l}{ 90\%} & \multicolumn{1}{l}{ 95\%} &       & \multicolumn{1}{l}{ 90\%} & \multicolumn{1}{l}{ 95\%} &       & \multicolumn{1}{l}{ 90\%} & \multicolumn{1}{l}{ 95\%} \\
\cmidrule{2-3}\cmidrule{5-6}\cmidrule{8-9}\cmidrule{11-12}    VEXP  & 37.3  & 38.7  &       & 13.3  & 13.5  &       & 10.1  & 10.4  &       & 6.9   & 6.4 \\
    CVEXP & 37.3  & 38.7  &       & 50.6  & 52.1  &       & 60.7  & 62.5  &       & 67.6  & 68.9 \\
    RCVEXP & 92.9  & 96.2  &       & 93.9  & 96.7  &       & 94.1  & 97.0    &       & 95.3  & 97.3 \\
    Card  & 5     & 7     &       & 5     & 6     &       & 5     & 6     &       & 8     & 7 \\
    Min \%Cont & 16.0 & 11.5 &       & 13.6 & 11.8 &       & 13.3 & 12.2 &       & 9.5 & 6.5 \\
    \bottomrule
    \end{tabular}%
}
  \label{tab:fers_loads}%
\end{table}%
\endgroup

\begingroup\renewcommand{\arraystretch}{0.6}
\begin{table}[H]
  \centering
  \caption{Correlation between the USPCs and the corresponding PCs for the Students ability dataset.}
  {\scriptsize
    \begin{tabular}{lrrrr}
    \toprule
          & \multicolumn{1}{l}{Comp 1} & \multicolumn{1}{l}{Comp 2} & \multicolumn{1}{l}{Comp 3} & \multicolumn{1}{l}{Comp 4} \\
\cmidrule{2-5}    USPCA 90\% & 0.96  & 0.97  & 0.94  & 0.79 \\
    USPCA 95\% & 0.98  & 0.98  & 0.98  & 0.94 \\
    \bottomrule
    \end{tabular}%
}
  \label{tab:fer_corpcs}%
\end{table}%
\endgroup
\subsubsection{SPCs with low cardinality}
\begin{figure}[H]
\centering
\includegraphics[width = 0.75\textwidth]
 {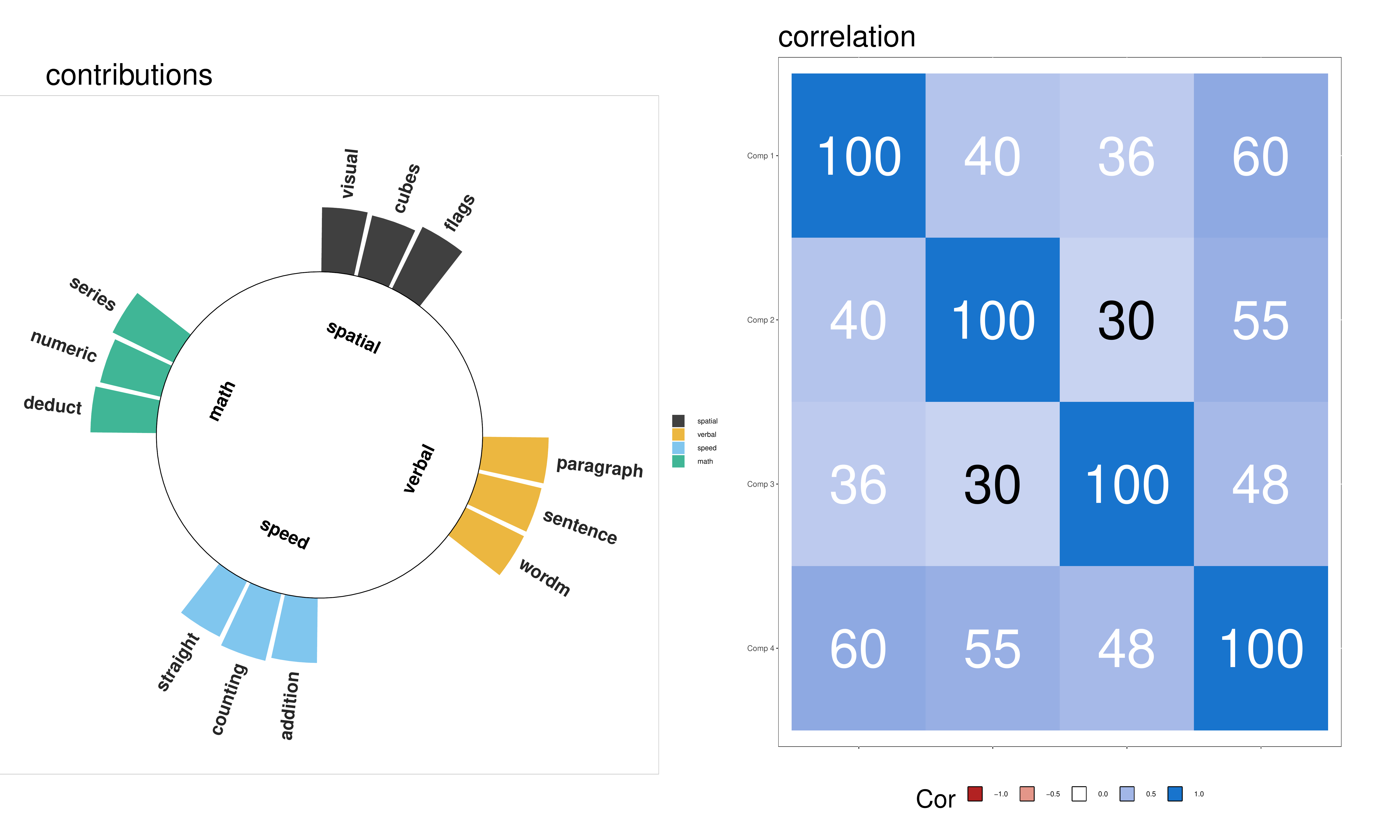}
\caption{Contributions of the first four PCs computed separately for each ability type and the their pairwise correlations of the resulting SPCs for the Students Ability dataset.}%
\label{fig:Fig_fermax3Circ}
\end{figure}

\cite[][Table 6]{fer} applied a new SPCA method (PDPCA) to this dataset obtaining, exactly , the PCs of the tests scores in the different abilities. These are simply averages of the variables  that we knew a--priori to measure the same ability, as shown in the plot on the left of Figure \ref{fig:Fig_fermax3Circ}. The corresponding summary statistics are shown in Table \ref{tab:fer_block}. The resulting SPCs are highly mutually correlated, as shown in the plot on the right of Figure \ref{fig:Fig_fermax3Circ}, and erratically with the PCs, as shown in Table \ref{tab:fer_block}.

I am not sure how these components can be useful. If such components are desired, they can be obtained without sophisticated algorithms. Another possibility os to apply LS SPCA to each Ability group, but on such a inconsistent dataset, the results would still be the PCs of each group.

%

\begingroup\renewcommand{\arraystretch}{0.6}
\begin{table}[H]
  \centering
  \caption{The first four columns show the summary statistics for the first four PCs computed separately for each ability  for the Students Ability dataset. The last four columns show the (percentage) correlation between these components and the first four PCs.}
    {\scriptsize
\begin{tabular}{lrrrrrlrrrr}
\cmidrule{1-5}\cmidrule{7-11}      & \multicolumn{1}{l}{\textbf{Comp 1}} & \multicolumn{1}{l}{\textbf{Comp 2}} & \multicolumn{1}{l}{\textbf{Comp 3}} & \multicolumn{1}{l}{\textbf{Comp 4}} &       &       & \multicolumn{4}{c}{\% Correlations} \\
\cmidrule{2-5}\cmidrule{8-11}VEXP  & 26.8  & 19.6  & 15.6  & 7.8   &       & PC    & \multicolumn{1}{l}{\textbf{Comp 1}} & \multicolumn{1}{l}{\textbf{Comp 2}} & \multicolumn{1}{l}{\textbf{Comp 3}} & \multicolumn{1}{l}{\textbf{Comp 4}} \\
\cmidrule{8-11}CVEXP & 26.8  & 46.3  & 61.9  & 69.7  &       & \textbf{PC1} & 75    & 78    & 66    & 86 \\
RCVEXP & 66.6  & 86    & 96.1  & 98.3  &       & \textbf{PC2} & 4     & -51   & 65    & 3 \\
Card  & 3     & 3     & 3     & 3     &       & \textbf{PC3} & 60    & -32   & -32   & 11 \\
 &  &  &  &  &       & \textbf{PC4} & 15    & 9     & 10    & -39 \\
\cmidrule{1-5}\cmidrule{7-11}\end{tabular}%
}
  \label{tab:fer_block}%
\end{table}%
\endgroup

As a final example, i show the results of running LS SPCA requiring that the cardinality of each component is equal to three. I do not recommend to constrain the cardinality a priori, rather then the variance explained, because it is impossible to foresee the effects of changing the cardinality.

\begin{figure}[H]
\centering
\includegraphics[width = 1\textwidth]
 {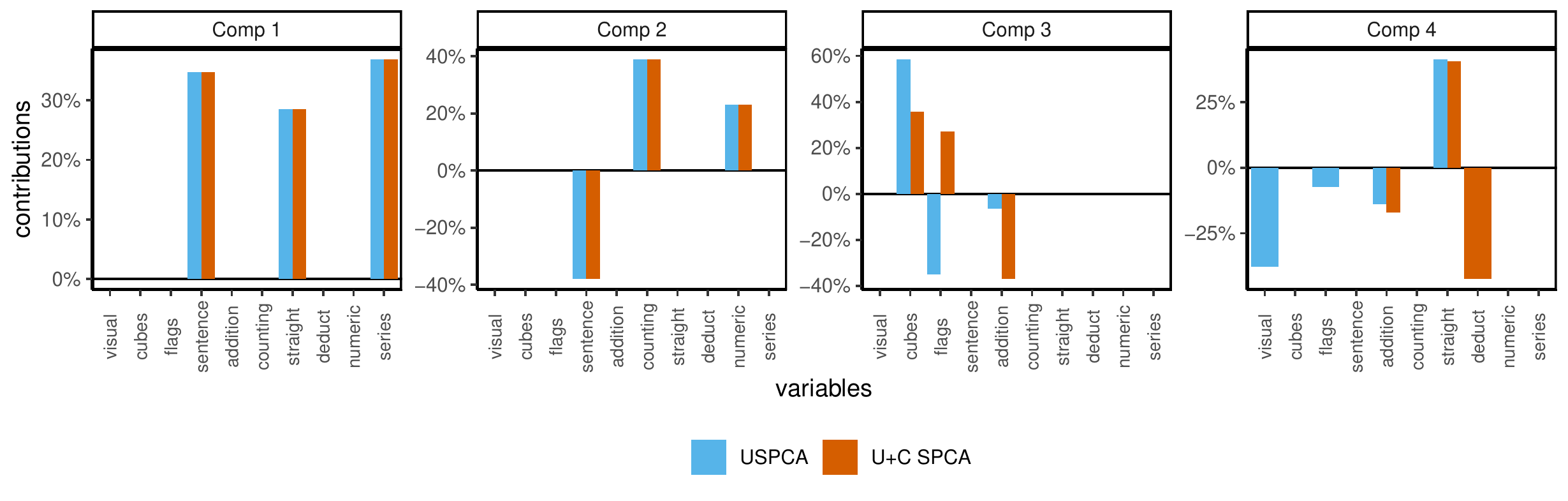}
\caption{Contributions of the first four USPCSs and of the LSSPCs computed requiring that only the first two are orthogonal. All SPCs are constrained to have cardinality at most three except the fourth USPC, which needs to have cardinality four.}%
\label{fig:Fig_ferCorPCmax3}
\end{figure}

Figure \ref{fig:Fig_ferCorPCmax3} shows the contributions of the USPCs constrained to have cardinality three for the first three components and four for the last (as required by orthogonality) together with the corresponding set of SPCs, all of cardinality three, the last two of which are not required to be orthogonal (CSPCs). The summary statistics are shown in Table \ref{tab:Fer:mixed_lsspca}.
In this case, there is a substantial improvement in efficiency by removing the orthogonality constraint. The resulting SPCs are mutually correlated but keep a substantial correlation with the PCs, as shown in Table  \ref{tab:fer_cor_SPCCard3}.

\begingroup\renewcommand{\arraystretch}{0.6}
\begin{table}[H]
  \centering
  \caption{Summary statistics for the first four USPCSs and of the LSSPCs computed requiring that only the first two are orthogonal. All SPCs are constrained to have cardinality at most three except the fourth USPC, which needs to have cardinality four.}
  {\scriptsize
    \begin{tabular}{lrrrrrrrrrrr}
    \toprule
          & \multicolumn{2}{c}{\textbf{Comp 1}} &       & \multicolumn{2}{c}{\textbf{Comp 2}} &       & \multicolumn{2}{c}{\textbf{Comp 3}} &       & \multicolumn{2}{c}{\textbf{Comp 4}} \\
\cmidrule{2-12}          & \multicolumn{1}{l}{USPCA} & \multicolumn{1}{l}{Mixed} &       & \multicolumn{1}{l}{USPCA} & \multicolumn{1}{l}{Mixed} &       & \multicolumn{1}{l}{USPCA} & \multicolumn{1}{l}{Mixed} &       & \multicolumn{1}{l}{USPCA} & \multicolumn{1}{l}{Mixed} \\
\cmidrule{2-3}\cmidrule{5-6}\cmidrule{8-9}\cmidrule{11-12}    VEXP  & 34.6  & 34.6  &       & 12.0    & 12.0    &       & 5.9   & 10.1  &       & 5.9   & 7.2 \\
    CVEXP & 34.6  & 34.6  &       & 46.6  & 46.6  &       & 52.4  & 56.7  &       & 58.3  & 63.9 \\
    RCVEXP & 86.0    & 86.0    &       & 86.4  & 86.4  &       & 81.3  & 87.9  &       & 82.3  & 90.1 \\
    Card  & 3     & 3     &       & 3     & 3     &       & 3     & 3     &       & 4     & 3 \\
    \bottomrule
    \end{tabular}%
    }
  \label{tab:Fer:mixed_lsspca}%
\end{table}%

\begin{table}[H]
  \centering
  \caption{The first four columns show the mutual correlations among the the first four LSSPCs computed requiring that only the first two are orthogonal. All SPCs are constrained to have cardinality at most three except the fourth USPC, which needs to have cardinality four. The last four columns show the correlations between the SPCs and the PCs.}
 {\scriptsize
    \begin{tabular}{lrrrrrrrrr}
    \toprule
          & \multicolumn{1}{l}{SPC 1} & \multicolumn{1}{l}{SPC 2} & \multicolumn{1}{l}{SPC 3} & \multicolumn{1}{l}{SPC 4} &       & \multicolumn{1}{l}{PC 1} & \multicolumn{1}{l}{PC 2} & \multicolumn{1}{l}{PC 3} & \multicolumn{1}{l}{PC 4} \\
\cmidrule{2-5}\cmidrule{7-10}    SPC 1 & 1     & 0     & 0.16  & -0.02 &       & 0.92  & -0.04 & -0.12 & 0.11 \\
    SPC 2 & 0     & 1     & -0.11 & 0.08  &       & 0.16  & 0.86  & 0.14  & -0.18 \\
    SPC 3 & 0.16  & -0.11 & 1     & 0.04  &       & 0.27  & -0.31 & 0.85  & 0.04 \\
    SPC 4 & -0.02 & 0.08  & 0.04  & 1     &       & -0.18 & 0.29  & 0.09  & 0.77 \\
    \bottomrule
    \end{tabular}%
}
  \label{tab:fer_cor_SPCCard3}%
\end{table}%
\endgroup

%
%
%
%
%
\section{Computational details}\label{sec:compDet}
I use the same notation used in Section \ref{sec:outline},which is:
$n\times p$ matrix $\bX$ is the data matrix with the columns centered to zero mean and
$\bS \propto \bX\trasp\bX$ is the covariance matrix. $\dX_j$ denotes a generic subset of variables and $\dSj \propto \dXjT\dXj$ its covariance matrix.
The SPCs are defined by $\bt_j = \dX_j\da_j$, where $\da_j$ is the vector containing only the nonzero loadings. In some cases the SPCs are expressed as combinations of all the variables as $\bt_j = \bX\ba_j$, where $\ba_j$ is the $p$-vector obtained by replacing the values missing in $\daj$ with zeroes.
\subsection*{USPCA: uncorrelated LS SPCA}
The loadings of the first USPC are computed as the generalized eigenvector satisfying
\begin{equation}\label{eq:uspc1}
  \dX\traspd{1}\bX\bX\trasp\dX_1\da_1 = \bdS_1\da_1\gamma_{max},
\end{equation}
where $\gamma_{max}$ is the largest generalized eigenvalue.
Given a set of $(j -1)$, $j \geq 2$, UPCs, $\bT = \bX\bA$, say, the constraints on the next USPC, $\btj = \bX\baj =
\dXj\daj$, are $\bT\trasp\btj = \bA\trasp\bS\baj = \bR\daj = \mathbf{0}$, where $\bR_j= \bT\trasp\dXj$. Let,
$\bC_j = \bR\trasps{j}\bigg(\bR_j\dSjinv\bR\trasps{j}\bigg)^{-1}\bR_j\dSjinv$, then the loadings of the $j$-th USPC satisfy\footnote{Note that $\dSjinv\bC_j = \bC\trasps{j}\dSjinv$ and that $\bC\trasps{j}\bC\trasps{j} = \bC\trasps{j}$. Since $\da_j$ must be in the span of $\bC\trasps{j}$, then $\bC\trasps{j}\da_j = \da_j$. Then, with some manipulation, this result can be obtained from the results in \cite{mer}.}
\begin{equation*}
  \bC_j\dX\traspd{j}\bX\bX\trasp\dX_j\bC\trasps{j}\da_j = \dSj\da_j\gamma_{max},\, j > 1,
\end{equation*}
Only for USPCA, $\gamma_{max}$ is equal to the variance explained by the component.
\subsection*{CSPCA: correlated LS SPCA}
Let $\bQ_{j + 1} =  \bQ_{j} - \bt_j\bt\trasps{j}/\big(\bt\trasps{j}\bt_j\big)$ be the residuals of $\bX$ orthogonal to the first $j$ CSPCs, with $\bQ_1 = \bX$. Then the loadings of the $j$-th CSPC satisfy
\begin{equation}\label{eq:cspc1}
  \dX\traspd{j}\bQ_j\bQ\trasps{j}\dX_j\da_j = \dSj\da_j\lambda_{max}.
\end{equation}
The first CSPC is equal to the first USPC.
\subsection*{PSPCA: projection LS SPCA}
The loadings of the $j$--th PSPC are obtained as the least squares estimates of the coefficients of the regression model
\begin{equation}\label{eq:mod_pspca}
  \br_j = \dXj\daj + \be_j,
\end{equation}
where $\br_j$ is the first PC of the residual matrix $\bQ_j$ defined for the CSPCs. A suitable subset $\dXj$ can be obtained with a variable selection algorithm. This subset can be then used to compute USPCs or CSPCs.
\subsection*{crude PSPCA: simple projection of the PCs}
In its simplest form, the PSPCs can be computed by regressing each PC $\bp_j$ onto a subset of variables, $\dXj$.
\subsection*{Variable selection}
Variables can be selected in different ways. A computationally efficient method is to use a regression variable selection algorithm on Equations \ref{eq:mod_pspca}. Otherwise, a backward elimination and a branch and bound algorithms are suggested in \cite{mer}.
\subsection*{Variance explained}
The variance explained by an SPC is simply the net variance of the projection of the data matrix onto it. Many computer packages offer an ANOVA function which will provide the extra sums of squares (sometimes called \emph{sequential}) for the regression of the $\bX$ matrix on the SPCs.

Otherwise, the variances explained can be computed manually.
If we let $\widehat{\bX}_j$ be the fitted values of the regression of $\bX$ onto the first $j$ SPCs, $\bt_j, \ldots, \bt_j$, the cumulative variance explained by these SPCs is equal to the sum of the squared elements of $SSR_j = \widehat{\bX}_j$ and the net variance explained by $\bt_j$ is equal to  $SSR_j - SSR_{j -1}$. More computationally efficient methods are: compute the cumulative variance explained when computing the orthogonal residuals; or apply a Householder decomposition to the SPCs and then compute the variance explained as for orthogonal components on this.
\subsection{Computational steps}
Algorithm \ref{algo:lsspca} describes the steps necessary for computing LS SPCA.

\begingroup
\renewcommand{\baselinestretch}{0.9}
\begin{algorithm}[H]\caption{generic algorithm for LS SPCA}%
\footnotesize{
    \begin{algorithmic}[1]
    \Procedure {lsspca}{$\bX,\, \alpha\in(0, 1),\, method = \{LSSPCA, PSPCA\}$}
    \State \textbf{initialize}
    \State {\hspace{1em}$\bQ_1 \gets \bX; \hspace{1em}j \gets 0; \hspace{1em}stopCompute \gets FALSE$}
    \State \textbf{end initialize}
    \While{(stopCompute = FALSE)}\Comment{\textbf{start components computation}}
        \State{$j \gets j + 1$}
        \State{$\bQ\trasps{j}\bQ_j\bw_j = \bw_j\lambda_{max};\, \brj = \bQj\bwj $} \Comment{compute first
        PC of $\bQj$}\label{algo:pwm}
        \State{$\dXj,\,\daj \gets \text{arg min}:\, ||\bX - \dXj\daj||^2 \leq
        \alpha$} \Comment{regression variable selection}\label{algo:fwdselect}
        \If{(method = LSSPCA)}\Comment{LS SPCA needs a function for it}
            \State{$\daj \gets LSSPCA(\dXj)$}
        \EndIf
        \State{$ \btj \gets \dX_j\daj$} \Comment{$j$-th sparse component}
        \State{Evaluate($stopRule) \gets \{TRUE, FALSE\}$}
        \If{(stopRule = FALSE)}
            \State{$ \bQ_{j+1} \gets \bQj - \frac{\btj\btjT}{\btjT\btj}\bQj$}
            \Comment{deflate $\bX$ of current component}\label{algo:deflX}
            \State{$cvexp(j) \gets sum\bigl((\bX - \bQ_{j+1})^2\bigr)$}
            \Comment{cumulative \vexp}\label{algo:cvexp}
        \Else
            \State{stopCompute $\gets TRUE$}\Comment{\textbf{terminate components
            computation}}
        \EndIf
    \EndWhile%
    \EndProcedure
    \end{algorithmic}}\label{algo:lsspca}
\end{algorithm}%
\endgroup
\subsection*{R package}
A lightweight \textsf{R} package is available for download on \textsf{Github} at \url{https://github.com/merolagio/LSSPCA/}. Otherwise it can be installed directly with the command
 \verb|devtools::install_github(merolagio/LSSPCA/)|, if the package \textsf{devtools} is available. The data used for the examples are included in the package with instruction for reproducing them.
\section{Concluding remarks}
LS SPCA does the remarkable job of sparsifying the PCs maintaining their original optimality. The SPCs are usually easier to interpret and to visualize than the PCs.

LS SPCA has several advantages over other SPCA methods. One is that it is transparent and can be easily implemented with standard statistical software. Different sets of SPCs can be computed and compared by modifying the minimal variance to be explained requirement or the algorithm used for variable selection.

Another advantage of LS SPCA is that it produces close approximations to the PCs and does not prefer correlated variables, which do not help to explain the variance of the data and make the cardinality unnecessarily high.

LS SPCA can be a useful tool for simplifying the interpretation of the  PCs. The \textsf{R} package with functions for computing and visualizing LS SPCA should be enough for experimenting with LS SPCA.
\bibliographystyle{apalike}
\bibliography{tut_SPCA}
\end{document}